\documentclass[useAMS,usenatbib,usegraphicx]{mn2e}

\newcommand{\aap}{A\&A}
\newcommand{\apj}{ApJ}
\newcommand{\apjl}{ApJ}
\newcommand{\apjs}{ApJS}
\newcommand{\apss}{Ap\&SS}
\newcommand{\araa}{ARAA}
\newcommand{\BP}{Ballesteros-Paredes}

\newcommand{\kms}{{\rm ~km~s}^{-1}}
\newcommand{\Lc}{L_{\rm c}}
\newcommand{\LJ}{L_{\rm J}}

\newcommand{\Mdense}{M_{\rm dense}}
\newcommand{\mh}{m_{\rm H}}
\newcommand{\Minf}{{\cal M}_{\rm inf}}
\newcommand{\mnras}{MNRAS}

\newcommand{\Msinks}{M_{\rm sinks}}
\newcommand{\Msinkdot}{\dot M_{\rm sinks}}

\newcommand{\Msun}{M_\odot}
\newcommand{\Msup}{M_{\rm sup}}
\newcommand{\mucrit}{\mu_{\rm crit}}

\newcommand{\nad}{n_{\rm AD}}
\newcommand{\nat}{Nature}

\newcommand{\nsink}{n_{\rm sink}}
\newcommand{\pasj}{PASJ}
\newcommand{\pasp}{PASP}
\newcommand{\pcc}{{\rm ~cm}^{-3}}
\newcommand{\psc}{{\rm ~cm}^{-2}}

\newcommand{\tff}{t_{\rm ff}}

\newcommand{\vinf}{v_{\rm inf}}
\newcommand{\VS}{V\'azquez-Semadeni}

\newcommand{\beq}{\begin{equation}}
\newcommand{\eeq}{\end{equation}}
\newcommand{\bea}{\begin{eqnarray}}
\newcommand{\eea}{\end{eqnarray}}
\newcommand{\bef}{\begin{figure}}
\newcommand{\eef}{\end{figure}}

\newcommand{\G}{\mbox{G}}

\newcommand{\pc}{\mbox{pc}}

\title[Molecular Cloud Evolution IV: Magnetic Fields]{Molecular Cloud
Evolution IV: Magnetic Fields, Ambipolar Diffusion, and the Star
Formation Efficiency} 
\author[\VS\ et al.] {Enrique \VS$^{1}$\thanks{E-mail:
e.vazquez@crya.unam.mx}, Robi Banerjee$^2$\thanks{E-mail:
banerjee@ita.uni-heidelberg.de}, Gilberto
C. G\'omez$^1$\thanks{E-mail:g.gomez@crya.unam.mx},  \and
Patrick Hennebelle$^3$\thanks{patrick.hennebelle@lra.ens.fr},
Dennis Duffin$^4$\thanks{duffindf@mcmaster.ca}, and Ralf
S. Klessen$^2$\thanks{E-mail:rklessen@ita.uni-heidelberg.de} \\  
$^{1}$Centro de Radioastronom\'\i a y Astrof\'\i sica,
Universidad Nacional Aut\'onoma de M\'exico, Campus Morelia,
Apdo. Postal 3-72,\\
Morelia, 58089, M\'exico\\
$^2$Zentrum f\"ur Astronomie der Universit\"at Heidelberg,
Institut f\"ur Theoretische Astrophysik, 69120 Heidelberg, Germany \\
$^3$Laboratoire de radioastronomie millim\'etrique (UMR 8112 CNRS), 
  \'Ecole Normale Sup\'erieure et Observatoire de Paris, 24 rue Lhomond,
  75231 Paris Cedex 05, France\\
$^4$Department of Physics and Astronomy, McMaster University, Hamilton,
Ontario L8S 4M1, Canada}

\begin{document}



\maketitle

\label{firstpage}

\begin{abstract}
We investigate the formation and evolution of giant molecular clouds
(GMCs) by the collision of convergent warm neutral medium (WNM) streams
in the interstellar medium, in the presence of magnetic fields and
ambipolar diffusion (AD), focusing on the evolution of the star
formation rate (SFR) and efficiency (SFE), as well as of the
mass-to-magnetic-flux ratio (M2FR) in the forming clouds. We find that:
1) Clouds formed by 
supercritical inflow streams proceed directly to collapse, while clouds
formed by subcritical streams first contract and then re-expand,
oscillating on the scale of tens of Myr. 2) Our suite of
simulations with initial magnetic field strength of $2$, $3$, and $4 \,
\mu\G$ show that only supercritical or marginal critical streams lead to
reasonable star forming rates. This result is not altered by the
inclusion of ambipolar diffusion.
3) The GMC's M2FR is a generally increasing function of time, whose
growth rate depends on the details of how mass is added to the GMC from
the WNM. 4) The M2FR is a highly fluctuating function of position in the
clouds. This implies that a significant fraction of a cloud's mass may
remain magnetically supported, while SF occurs in the supercritical
regions that are not supported. 5) In our simulations, the SFE
approaches stationarity, because mass is added to the GMC at a similar
rate at which it converts mass to stars. In such an approximately
stationary regime, we find that the SFE provides a proxy of the
supercritical mass fraction in the cloud. 6) We observe the occurrence
of buoyancy of the low-M2FR regions within the
gravitationally-contracting GMCs, so that the latter naturally segregate
into a high-density, high-M2FR ``core'' and a low-density, low-M2FR
``envelope'', without the intervention of AD.

\end{abstract}
 

\section{Introduction} \label{sec:intro}

Magnetic fields in molecular clouds (MCs), and the gradual
redistribution of magnetic flux within them by ambipolar diffusion (AD),
have been thought to be crucial ingredients in regulating star formation (SF)
and its efficiency (SFE) for over two decades \citep[see, e.g., the
reviews by][and
references therein]{SAL87, Mousch91}. In early studies, MCs were
considered to generally have strongly magnetically subcritical
mass-to-flux ratios (M2FRs), and supercritical clouds were believed to be
rare \citep[e.g., ][sec. 2.1]{Mousch91} so that the time for AD to allow
their cores to become supercritical turned out to be very long, of order
10-20 times larger than the clouds' free-fall time ($\tff$)
\citep[e.g.,][]{CM94, BM94}. In this ``standard model'' of
magnetically-supported, AD-mediated subcritical clouds, the general
mechanism for the formation of low-mass stars was through the slow
gravitational contraction of isolated cores containing a very small
fraction of the clouds' mass, thus accounting for the very low observed
global SFE of giant MCs (GMCs) \citep{Myers_etal86, Evans_etal09}. 

However, subsequent studies have suggested that both MCs \citep{McKee89}
and their clumps \citep{BM92, MG88, Crutcher99, Bourke_etal01, CHT03, TC08}
are close to being magnetically critical, with a moderate preference for
being supercritical. This implies that, if a MC is subcritical, it is
expected to be only moderately so as well, in which case the time for
cores within it to become locally supercritical may be almost as short
as the cores' free-fall time \citep{CB01, VS_etal05}. Moreover, MCs are
generally believed to be supersonically turbulent \citep[see, e.g., the
reviews by][and references therein]{VS_etal00, MK04, ES04, BP_etal07,
MO07}, and in this case, AD must be treated nonlinearly. As a
consequence, its characteristic timescale is expected to decrease,
suggesting again that it may be comparable to the free-fall time
\citep{FA02, Heitsch_etal04}. Thus, the gravitational contraction and
collapse of a star-forming region must occur rapidly, essentially in the
timescales corresponding to those of a magnetically supercritical
region, which is of the order of a few free-fall times \citep{OGS99,
HMK01, VS_etal05, Galvan_etal07}, although the subcritical envelope may
still be held up by the magnetic field, since, at the envelopes' lower
typical densities, the AD timescale is indeed much longer than the
dynamical timescales.

Moreover, the realization that the majority of MCs may be supercritical,
and that most stars, including low-mass ones, form in cluster-forming
regions \citep{LL03}, have forced a reconsideration of the problem, to
accomodate the fact that the standard model's paradigm of low-mass star
formation in strongly subcritical clouds may be the exception rather
than the rule, even for the formation of low-mass stars.\footnote{One
instance of such an infrequent, strongly magnetized cloud may be the
Taurus MC \citep{Heyer_etal08}} Unfortunately, if most MCs are
supercritical, one must once again face the old \citet{ZP74} conundrum
that the Galactic star formation rate (SFR) should be roughly two orders
of magnitude larger than the one observed, of $\sim 3$--$4 \Msun$
yr$^{-1}$ \citep[see the supplementary material of][]{Diehl_etal06}.
This is because globally supercritical clouds cannot be supported by the
magnetic field, and thus should be collapsing as a whole.

Turbulence is often invoked as an additional source of support against
the clouds' self-gravity, as if it were simply an extra source of
pressure \citep{Chandra51}. Such a treatment, however, neglects the
fundamental property of turbulence that the largest velocities occur at
the largest scales, a property which is reflected in \citet{Kolmog41}'s
famous energy spectrum of the turbulence. Studies taking this into
account have generally only considered it from the point of
view of the energetics involved \citep[e.g.,][]{Bonaz_etal87, VG95}, but
have neglected the vector nature of the velocity field. Having the
largest velocity differences at the largest scales implies that the
effect of classical vortical turbulence within a cloud or clump should
be primarily to distort it, rather than support it
\citep{BP_etal99a}. In highly supersonic turbulence,
\citet{VS_etal08} have recently noted that the non-thermal motions
within clumps must contain a significant compressive component, even if
the driving is purely solenoidal. This compresive component,
rather than opposing gravity, aids it or is driven by it. This is
consistent with recent suggestions that, for example, the Orion A cloud
is collapsing, and producing the Orion Nebula Cluster (ONC) in the
process \citep{HaBu07}, that the non-thermal motions
within the clump NGC 2264-C correspond mainly to gravitational
contraction, rather than to isotropic turbulence \citep{PHA07}, that
massive star-forming regions may be immersed in large-scale accretion
flows \citep{Galvan_etal09, Csengeri_etal10, Schneider_etal10}, and that
MCs in the Large Magellanic Cloud seem to follow an evolutionary trend
such that more evolved clouds are more massive \citep{Fukui_etal09}. Also, if
the non-thermal motions within clouds were homogeneous and isotropic
turbulence, it would be difficult to understand the common observation
that SF occurs at localized spots within the clouds, rather than scattered
throughout their volumes \citep[e.g., ][]{Kirk_etal06, Evans_etal09}. 

These findings all suggest that an important, and perhaps even dominant,
component of the nonthermal motions observed in MCs and their
substructure is actually converging flows, which may be driven by
gravity or by external compressions. In fact, a
model in which the nonthermal motions in MCs are a
gravitationally-driven mass cascade has been recently proposed by
\citet{Field_etal08}. Moreover, the collision of converging flows has
been shown to produce turbulence in the compressed layers formed by them
\citep{Hunter_etal86, Vishniac94, WF00, Heitsch_etal05, Heitsch_etal06,
VS_etal06}. \citet{KH10} have recently shown that the turbulent kinetic energy
observed in objects as diverse as galactic disks, MCs, and protostellar
accretion disks is in general consistent with being driven by infall from
the environments of those objects. However, if the turbulence is being
driven by the gravitational contraction, it cannot be expected to halt
the contraction that drives it. Thus, one is faced with the
Zuckerman-Palmer conundrum again.

One way to avoid the Zuckerman-Palmer conundrum is if large chunks of
the molecular gas in the Galaxy are indeed magnetically subcritical and
thus supported against gravity, while SF occurs precisely in those
regions that are not, as recently suggested by \citet{Elm07}. Thus, the
chaotic spatial and statistical distributions of the physical variables,
produced by the turbulence in the forming MCs, may play
a key role in the control of the SFR and the SFE. 

In this paper we present numerical simulations of the formation and
evolution of MCs, starting from their formation out of
generic compressions in the warm neutral medium (WNM), and reaching up
to their star-forming epochs, in the presence of magnetic fields and AD,
in order to investigate the production of sub- and supercritical
regions, and the rate at which clouds with various environmental
conditions form stars. We focus on the effect of AD on the SFE and the
global evolution of the clouds. The plan of the paper is as follows: in
sec.\ \ref{sec:general} we present some general considerations on the
evolution of the M2FR in MCs. In sec.\ \ref{sec:num_mod} we
present the numerical model and the parameters of the simulations. In
sec.\ \ref{sec:results} we then present our results on the variability of the
M2FR, the evolution of the SFE, and the role of AD in the evolution
of both sub- and supercritical clouds. Finally, sec.\
\ref{sec:conclusion} presents a summary and our conclusions.

\section{Evolution of the M2FR in molecular clouds: from sub- to
supercritical} \label{sec:general} 

Although the most commonly considered mechanism for increasing the M2FR
of a certain density enhancement is the redistribution of magnetic flux
among the central flux tubes of a cloud by AD \citep{MS56, Mousch77},
another important, yet often-neglected mechanism is that, for a uniform
medium permeated by a given mean magnetic field strength $B_0$, there is
always a certain length along the field (termed the ``accumulation
length'' by Mestel 1985; see also Shu et al. 2007) such that flux tubes
longer than that contain enough mass per unit area to be magnetically
supercritical.  The criticality condition in terms of the mass column
density $\Sigma = \rho L$ and the field strength $B_0$ for a cylindrical
geometry is \citep{NN78},
\beq
\Sigma/B_0 \approx (4 \pi^2 G)^{-1/2}  \approx 0.159 \, G^{-1/2},
\label{eq:NN78}
\eeq
where $\rho$ is the mass density and $L$ is the cylinder length.  This
condition gives the accumulation length, in terms of fiducial values
representative of the ISM in the solar neighborhood, as \citep{HBB01}
\begin{equation}
\Lc \approx 470 \left(\frac{B_0}{5 \mu{\rm G}} \right) \left(\frac{n}{1
\pcc} \right)^{-1}~{\rm pc},
\label{eq:acc_length}
\end{equation}
where $n = \rho/(\mu \mh)$ is the number density of the medium, $\mh$ is
the Hydrogen mass, and $\mu$ is mean particle weight, taken as $\mu =
1.27$. In principle, if the Galactic field is primarily azimuthal, then
the Galactic ISM at large is magnetically supercritical in general,
because field lines do not end, and thus sufficiently long distances are
always available along them.\footnote{Note, however, that
supercriticality does not necessarily imply collapse, since the gas may
be thermally or otherwise supported, as is likely the case for the
diffuse warm medium at scales of hundreds of parsecs.} Thus, {\it the
M2FR of a system is not a
uniquely defined, absolute parameter, but 
rather depends on where the system's boundaries are drawn.} We also
stress that the M2FR depends on the local geometry of the considered
system. For instance, a system with spherical symmetry has a slightly
lower critical value of $\mucrit \approx 0.13 \, \G^{-1/2}$
\citep{Mouschovias76}. Measuring the criticality of the streams with
respect to this value would lead to a supercritical configuration for our
runs B3 rather than sub-critical configurations.

Now consider a cloud or clump that is formed by the accumulation of gas
along field lines in general.\footnote{Since compressions perpendicular
to the magnetic field cannot induce collapse, and compressions oblique
to the field can produce collapse by reorienting the directions of the
flow and the field lines \citep{HP00}, our assumed configuration
involves no loss of generality.} In the rest of this discussion, we will
generically refer to the resulting density enhancement as a ``cloud'',
referring to either a cloud, a clump, or a core. Although redistribution
of matter along field lines does not in principle affect the {\it total}
M2FR along the full ``length'' of a flux tube, this length is a rather
meaningless notion, since the flux tube may extend out to arbitrarily
long distances. What is more meaningful is the M2FR {\it of the dense
gas that makes up the cloud}, since the cloud is denser than its
surroundings, and thus it is the main source of the self-gravity that
the field has to oppose. In fact, for the formation of a cloud out of
flow collisions in the WNM, the cloud's density is $\sim 100$ times
larger than that of the WNM \citep{HP99, KI02, Heitsch_etal05, AH05,
VS_etal06, Henneb_etal08, Banerjee_etal09}, and so the latter's
self-gravity is negligible. Thus, in 
this problem, natural boundaries for the system are provided by the
bounding surface of the dense gas, allowing a clear working definition
of the M2FR.


However, contrary to the very common assumption of a constant cloud
mass, the formation of clouds by converging gas streams implies that the
cloud's mass is a (generally increasing) function of time
\citep{BHV99, VS_etal07, VS_etal10, Banerjee_etal09, KH10}, a result
that has recently 
received observational support \citep{Fukui_etal09}. This means that,
{\it within the volume of 
the cloud, the M2FR is also an increasing quantity}, since the flux
remains constant if the flow is along field lines, while the mass
increases \citep[see also][]{Shu_etal07}. If the cloud starts from
essentially zero mass, this in turn implies that the M2FR of a cloud is
expected to start out strongly subcritical (when the cloud is only
beginning to appear), and to evolve towards larger values at later
times. Rewriting eq.\ (\ref{eq:acc_length}) for the column density, we
see that the cloud becomes supercritical when
\begin{equation}
N_{\rm cr} = 1.45 \times 10^{21} \left(\frac{B_0}{5 \mu {\rm G}}
\right) \psc,
\label{eq:Sigma_crit}
\end{equation}
where $N \equiv \Sigma/\mu m_H$ is the number column density along field
lines.

The critical column density for magnetic criticality given by eq.\
(\ref{eq:Sigma_crit}) turns out to be very similar, at least for solar
neighbourhood conditions, to the critical column density of hydrogen
atoms necessary for cold atomic gas to become molecular, $N_{\rm H}
\sim 1$--2$ \times 10^{21}\psc$ \citep[e.g.,][]{FC86, vDB88, vDB98,
HBB01, GM07a, GM07b, Glover_etal10}. Thus, {\it the evolution of a cloud
is such that it starts out 
as an atomic and subrcritical diffuse cloud \citep{VS_etal06} and,
as it continues to accrete mass from the warm atomic medium, it
later becomes molecular and supercritical, roughly at the same time}
\citep{HBB01}. This is fully
consistent with the observation that diffuse atomic clouds are in
general strongly subcritical \citep{HT05}, while MCs are approximately
critical or moderately supercritical \citep{Crutcher99, Bourke_etal01,
TC08}. 

Moreover, the critical column density given by eq.\
(\ref{eq:Sigma_crit}) is also very similar to that required for
rendering cold gas gravitationally unstable, which is estimated to be
\beq
N_{\rm grav} \approx 0.7 \times 10^{21} \left( \frac{P/k}
{3000~ K \pcc} \right)^{1/2}\psc
\label{eq:N_self-grav}
\eeq
\citep{FC86, HBB01}. Thus, at solar neighborhood conditions, a forming
cloud is expected to become molecular, magnetically supercritical, and
self-gravitating at roughly the same time.

It is important to note that the mass accretion onto a cloud due to gas
stream collisions is likely to start along essentially just one
dimension. This mode of mass accretion may be slow, and it has been
argued that it may involve excessively long times
\citep[e.g.,][]{MO07}. However, numerical simulations of the process
show that, once the gas has transitioned to the cold, dense phase, it
soon becomes gravitationally unstable, even though it may remain mainly
in the atomic phase, and three-dimensional gravitational contraction can
then ensue, providing a much faster mode for increasing the column
density \citep{VS_etal07, Elm07, HH08, Henneb_etal08,
Banerjee_etal09}. The same is true if the global 
convergence of the flow is driven by larger-scale gravitational
instabilities \citep[e.g., ][]{Kim_etal03, LMK05, KO07}.
Of course, this increase of the column density due to
gravitational contraction of the dense gas is only relevant to molecule
formation. During such a process, the M2FR remains constant if the
cloud's mass remains fixed or varies on timescales much longer than the
contraction, and the latter occurs under ideal MHD
conditions. The gravitational contraction can only contribute to a
further increase of the M2FR if the gravitational potential of the cloud
causes it to accrete further amounts of diffuse gas, which 
transitions to the dense phase as it is incorporated into the bulk of
the cloud.

In all of the processes discussed so far, AD has not played a role. This
is of course due to the well known fact that AD is not dynamically
relevant until densities as high as $\nad \sim 10^5 \pcc$ are reached
\citep{Mousch_etal85}. Such densities are only reached in the dense
cores of MCs and therefore AD is not expected to cause any important
effects on the global evolution of MCs. Such cores, however, may be
magnetically subcritical if they form by turbulent compressions within
the cloud before AD becomes locally important, even if the cloud is
globally supercritical. This is because, under ideal MHD, a core formed
within an initially uniform cloud must have a smaller M2FR than that of
the cloud \citep{VS_etal05}. In a sense, the core repeats the pattern
followed by its parent cloud, initially being strongly subcritical and
evolving towards higher values of the M2FR, being limited by the M2FR of
its parent cloud, until AD becomes important and allows its M2FR to
overtake that of the parent cloud, perhaps becoming supercritical and
allowing the core to collapse. However, this notion has not been tested
in the context of the global evolution of a GMC, in particular taking
into account the property that the cloud's M2FR should evolve (generally
increasing) with time. In the remainder of the paper we investigate this
scenario, by means of numerical simulations of the formation and
evolution of a GMC, including AD, and focusing in particular on the
resulting SFE. We are particularly interested in the star-forming
properties of the cloud as it transits from sub- to supercritical.


\section{The numerical model} \label{sec:num_mod}

\subsection{The numerical code and setup} \label{sec:code}

We use the adaptive mesh refinement (AMR) code FLASH
\citep{Fryxell_etal00} with MHD, modified to include the AD module
developed by \citet{DP08}. 
The AD treatment takes the single-fluid approximation, and uses a simple
prescription to avoid the need to track the ion density in this
approximation. This prescription essentially turns off AD at low ($n \la
10^3 \pcc$) densities. For more details, we refer the reader to
\citet{DP08}. 

The simulations also use a sink particle prescription
\citep{BBP95, Jappsen_etal05, Banerjee_etal09, Federrath_etal10}. A sink
particle is  
created in a cell if the density there reaches a threshold density
$\nsink = 2\times 10^5 \pcc$, and the cell is a local minimum of the
gravitational potential. When a cell forms a sink, the latter takes all
the mass in excess of $\nsink$ in the region where the density $n$
satisfies $n \ge \nsink$. The sink particles have an accretion radius of
0.065 pc, corresponding to roughly 1 Jeans length at $n_{\rm sink}$ and
$T \sim 20$ K.

Concerning the heating and cooling, we use the same prescription used in
\citet{VS_etal07} and \citet{Banerjee_etal09}, which is derived from the
fit by \citet{KI02} to the results of the chemistry and cooling
calculations of \citet{KI00}. This prescription implies that the
simulated ISM is thermally unstable in the density range $1 \la n \la 10
\pcc$, which, under thermal balance between heating and cooling,
corresponds to the temperature range $5000 \ga T \ga 500$ K. 

We model the convergence of WNM flows as the collision of two
large-scale cylindrical streams. Our setup is similar (though not
identical) to the non-magnetic SPH simulation of \cite{VS_etal07}
labeled L256$\Delta v0.17$ (see their Fig. 1). Each stream is 112 pc
long and has a radius of 32 pc. The streams collide at the plane $x=0$
pc, and are embedded in a $(256~{\rm pc})^3$ simulation box, in which
the coordinates range from $-128$ to 128 pc. The
numerical box is periodic, and the streams are completely contained
within it, ending at a distance of 16 pc from the $x$ boundaries. The
resulting cloud occupies a relatively small volume far from the
boundaries, and so it can interact freely with its diffuse environment,
with relatively little effect from the boundaries. Most importantly, the
cloud is free to grow by accretion from the WNM.

The cylindrical streams are given an initial, moderately supersonic
inflow velocity $\vinf$ so that they collide at the centre of the
numerical box. The inflow speed of each stream is measured with respect
to the isothermal sound speed of $5.7 \kms$ that corresponds to the
initial temperature of 5000 K. This isothermal inflow Mach number is
denoted $\Minf$. We also add
10\%  random velocity perturbations to
the bulk stream speeds, in order to trigger the instabilities that
generate turbulence in the forming cloud \citep{Vishniac94,
Heitsch_etal05, Pittard_etal05, VS_etal06}. The box is initially filled
with WNM at a uniform density of $n = 1 \pcc$ ($\rho = 2.12 \times
10^{-24}$ g$\pcc$, using a mean atomic weight of 1.27). At the
temperature of $T = 5000$ K for the warm phase, this implies that the
cold phase comes into hydrostatic thermal pressure balance with the WNM
at a density $n \approx 100 \pcc$
\citep[see Fig.\ 2 of ][]{VS_etal07}. However, in our simulations the
density of the cold phase is higher, because it is in balance with the
sum of the thermal and the ram pressure of the colliding streams
\citep{VS_etal06, Henneb_etal08, Banerjee_etal09}.

\subsection{Resolution issues} \label{sec:resolution}

We start our simulations at a base resolution of $512^3$,
i.e. $\Delta x = 0.5\, \pc$, at the convergence point of the
flows. Additonally, we allow the code to refine up to 4 additional
levels, the highest of which corresponds to a maximum resolution of
8192 grid points, or a grid spacing of $\Delta x = 0.03$ pc in each
direction.  For the dynamical mesh refinement we use a Jeans-type
criterion \citep[][see, however, Federrath et al. (2011, submitted) for
a more stringent criterion in the presence of magnetic fields]
{Truelove_etal97}, requiring the local Jeans length to 
be resolved with at least 10 grid cells while refining is active. Beyond
the last refinement level, the Jeans length begins to be more poorly
resolved, until a maximum allowed density of $\nsink = 2 \times 10^5
\pcc$ is reached, at which a sink particle is formed. We refer the
reader to \citet{Banerjee_etal09} for a discussion of the justification
and possible limitations of this choice with regards to thermal
issues. Here we discuss issues related with gravity and AD.

The value of $\nsink$ we use was
chosen in order to reasonably ensure that the material going into sink
particles is actually gravitationally bound. Indeed, 
\citet{Galvan_etal07} found that, when cores are defined by means of a
density threshold, most cores defined by thresholds $\ga 10^5 \pcc$
proceeed to collapse. Instead, when cores are defined by lower
thresholds (say, $\sim 10^4 \pcc)$, a significant fraction of them is
transient, rebounding instead of collapsing \citep[see also][]
{BP_etal03, Klessen_etal05, VS_etal05}. Now, at $n = \nsink$ and $T
= 20$ K, the Jeans length is $\LJ = 0.066$ pc, and so we marginally fail
to fulfill the minimum Jeans criterion, of resolving $\LJ$ with at least
4 cells. However, this should not introduce any
significant errors, as we are not concerned here with the fragmentation
of the core into multiple stars, nor with their mass distribution, but
only with the total mass going into stars.

Another issue is that numerical diffusion can have an effect similar to
that of AD, as discussed by \citet{KHM00}. Specifically, since the scale
of the densest cores is a few grid cells, numerical diffusion can cause
the magnetic flux to diffuse out of them, in a similar manner to
AD. Indeed, we do occasionally observe the occurrence of gravitational
collapse in magnetically subcritical, ideal MHD simulations, in which
theoretically this should not occur. Moreover, since $\nsink$
effectively constitutes an upper limit to the density that can be
reached by any cell in the simulation, and since we have chosen a value
of $\nsink$ that is of the same order as $\nad$, the highest densities
in the code will be of the order of $\nad$, and numerical diffusion and
AD will have comparable effects. This limitation could be avoided by
using an even larger number of refinement levels but, since the
simulations are already very numerically expensive ($\sim$ 200,000 CPU
hours per run), this option is not presently feasible. Alternatively, we
could give up on satisfying the Jeans criterion, simply raising $\nsink$
without increasing the allowed number of refinement levels, an option
that we may attempt elsewhere. In any case, the effect of numerical
diffusion will again be to allow SF to occur more readily than if
mediated by AD alone in subcritical cases, and so our SFRs must again be
considered upper limits to the ones caused by AD alone.

\subsection{The simulations} \label{sec:simulations}

We consider five numerical simulations with three reasonably realistic
values of the initial, uniform magnetic field $B_0$, of 2, 3, and 4
$\mu$G, respectively. These values span the observed range of values of
the uniform component of the Galactic magnetic field \citep{Beck01}. The
initial field is along the $x$-direction.

With respect to the cylindrical criticality criterion, eq.\
(\ref{eq:NN78}), these cases respectively correspond to $\mu/\mucrit
\approx 1.36$, 0.91 and 0.68, so that the first case is magnetically
supercritical while the other two are subcritical. Note, however, that
the subcritical cases are only so because of the finite extent (256 pc)
of the numerical box. For $B_0 = 3$ and $4 \mu$G, lengths of 280 and 380
pc, respectively, would be required to render the system magnetically
critical.  Also, because the critical value of the mass-to-flux
ratio depends on the geometry of the considered configuration,
individual (molecular) clumps could be supercritical if compared to the
slightly lower $\mucrit$ of \cite{Mouschovias76} (see discussion in
Sec.~\ref{sec:M2FR_distr}).

The supercritical case is considered only in the AD regime, as we do not
expect the absence of AD to make a significant difference in this
case. The $B_0 = 3$ and $4 \mu$G cases are considered both in the
``ideal'' and AD regimes, to investigate the effect of AD on the
star-forming properties of magnetically subcritical clouds. Note that we
have written the word ``ideal'' within quotation marks because we cannot
avoid the effect of numerical diffusion, even if we turn off the
AD. Except for the value of the magnetic field strength and whether AD
is on or off, the simulations are otherwise identical, all having an
inflow speed of $13.9 \kms$, corresponding to an isothermal Mach number
$\Minf = 2.44$ with respect to the unperturbed, initial medium at $T
= 5000$ K.

The runs are labeled mnemonically, so that the first two characters
of the run's name indicate the field strength in $\mu$G (e.g., ``B3''
denotes $B_0 = 3 \mu$G), and the last two denote whether the simulation is
in the ideal MHD case (``MH'') or includes AD (``AD''). Table
\ref{tab:run_params} summarizes the parameters used in each of the five
runs. The last column in this Table gives the maximum time reached
by each simulation. 

\begin{table}
\caption{\sc Run parameters}
\begin{tabular}{ccccc}
\hline
Run       & $B_0$      & AD   & $\mu/\mucrit$ & Final time\\
name      & [$\mu$G]   &      &               & [Myr]     \\
\hline
B2-AD     & 2.0      & on     &    1.36       & 31.4      \\
B3-MH     & 3.0      & off    &    0.91       & 26.1      \\
B3-AD     & 3.0      & on     &    0.91       & 35.6      \\
B4-MH     & 4.0      & off    &    0.68       & 48.4      \\
B4-AD     & 4.0      & on     &    0.68       & 59.2      \\
\hline
\label{tab:run_params}
\end{tabular}
\caption{Initial conditions and the final simulatation time of the
runs presented in this work. To calculate the magnetic critically of the
entire system we use the critical mass-to-flux ratio, $\mucrit \approx
0.16 \, \G^{-1/2}$, for a cylindrical geometry of \citet{NN78}} 
\end{table}

\subsection{Considerations on measuring the mass-to-flux ratio}
\label{sec:consid_M2FR}

In what follows, we will be presenting measurements of the M2FR in
various regions of the simulations. However, this is not an unambiguous
task in general, and in fact the M2FR can be measured using different
procedures. In principle, the M2FR should be measured along flux tubes, in
order for the measurement to be directly representative of the dynamical
effect of the field on the gas. Thus, the measurement should be
performed by tagging a bundle of field lines, and integrating the
density along the path defined by them. Unfortunately, such a
measurement is extremely difficult to perform, even in the
simulations. A magnetic flux tube may lose coherence if the field lines
that compose it diverge from each other at long distances. Also, the
field near and within the cloud can be significantly distorted, due both
to the turbulence in the cloud, and to its gravitational contraction,
even if the initial flow direction is along the field lines. 

Observationally, the M2FR is often estimated by measuring the ratio of
column density to magnetic field strength, $N/B$, along lines of sight
(LOSs) through the cloud of interest \citep[e.g.][]{CHT03}. However,
this procedure actually intersects many different flux tubes, and thus
gives only an approximation to the actual M2FR of a single flux tube.
As discussed by \citet{Crutcher99}, if a cloud is flattened, its plane
is perpendicular to the magnetic field lines, and the system is observed
at an angle $\theta$, then the observed M2FR, or equivalently, the $N/B$
ratio, will be related to the actual one by $(N/B)_{\rm obs} = N/(B
\cos^2 \theta)$. Alternatively, as is the case for the measurements we
present below, if the observation is performed along an LOS that is
perpendicular to the cloud, but the magnetic field is at an angle
$\theta$ with respect to the LOS and to the normal to the plane, then we
have that $(N/B)_{\rm obs} = N/(B \cos \theta)$, with $|\theta| \le
\pi/2$. In either case, $(N/B)_{\rm obs}$ tends to overestimate the
actual value, and on some occasions very large values may be
artificially measured. This implies that a map of $(N/B)_{\rm obs}$ is
actually a map of {\it upper limits} to the actual $N/B$. This led
\citet{Crutcher99} to introduce statistical correction factors of
1/2--1/3 to the set of M2FR values obatined in the observations he
considered. 

The equivalent procedure for the simulation data is to measure M2FR
along LOSs through the clouds in our simulations. We choose the LOSs to
lie along the $x$-direction, since this is the direction of the mean
magnetic field and of the colliding WNM streams, and thus it is the
direction along which the column density is dynamically relevant.
Specifically, we then measure the M2FR as
\beq
\mu = \frac{\Sigma}{\langle B_x \rangle} \equiv \frac{\int_L \rho
dx}{L^{-1}\int_L B_x dx},
\label{eq:proj_mu}
\eeq
where $L$ is the stretch of the cloud along the $x$ direction. The path
$L$ is chosen so as to contain the full extent of the cloud's
thickness. This is done for every position on the plane of the cloud, to
obtain maps of the M2FR over the cloud's surface. We refer to this as
``the projection method'' of measuring the M2FR in the simulations.

\begin{figure*}
\includegraphics[width=0.45\hsize]{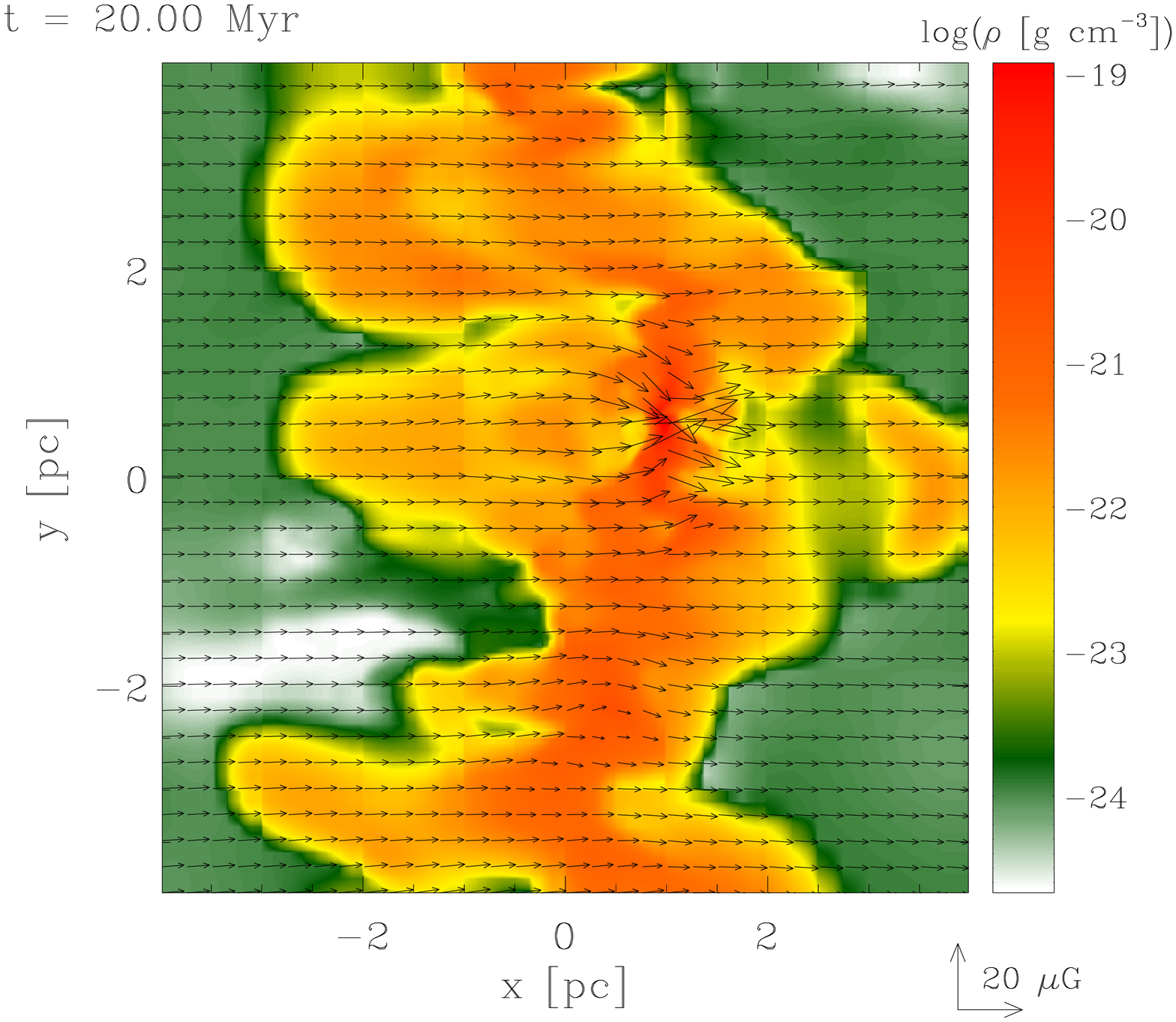}
\includegraphics[width=0.45\hsize]{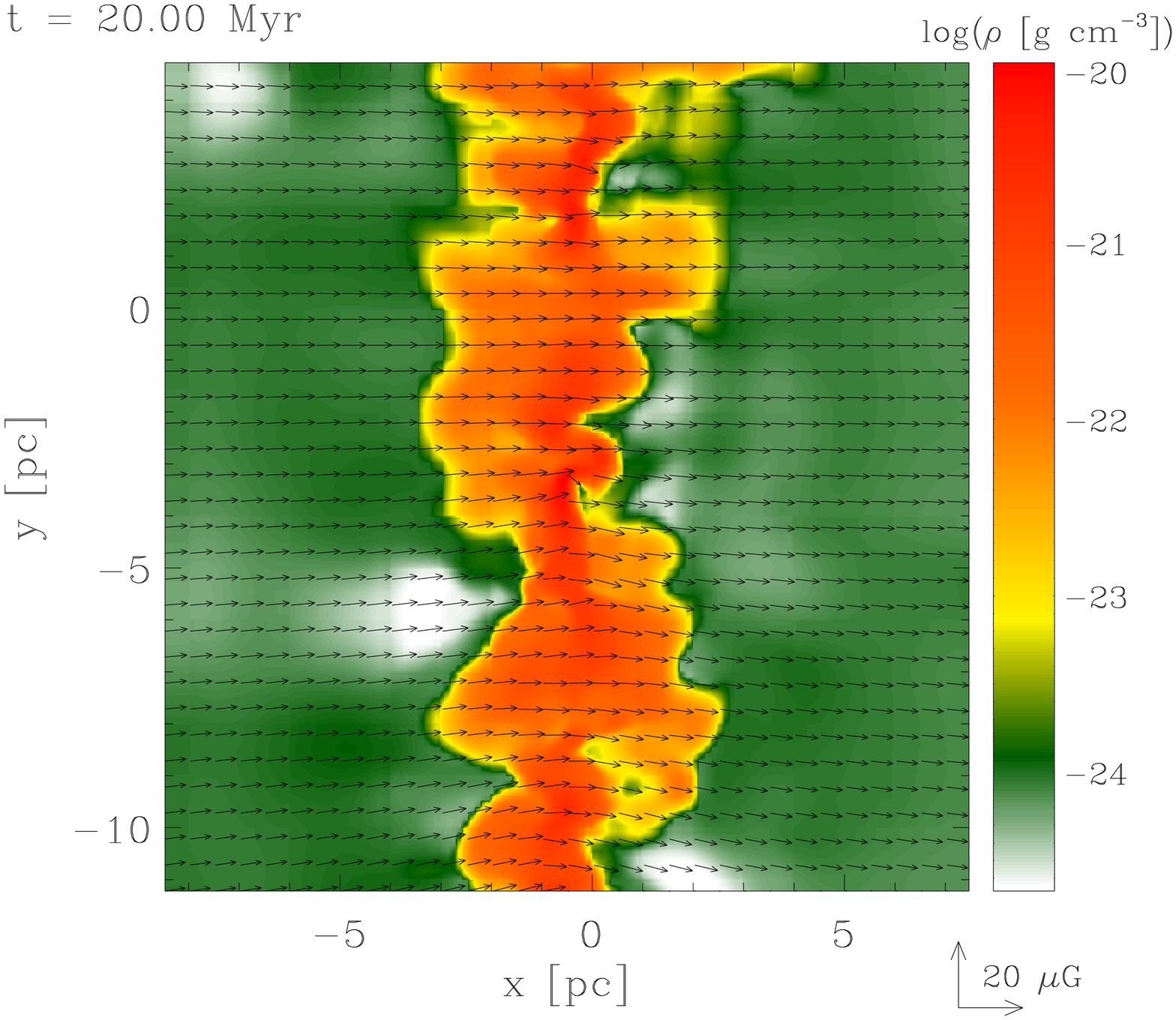}
\caption{Cross section through the simulation centre on the $(x,y)$
plane, showing the density field on the plane and the component of the
magnetic field on this plane, for the B3-MH ({\it left panel}) and for
the B4-MH ({\it right panel}) runs at $t=20$ Myr. The initial flow
collision occurred along the $x$ direction, which is also the direction
of the mean magnetic field.}
\label{fig:dens_mag_xy}
\end{figure*}

In order to appreciate the amount of distortion that may be present in
the field within the clouds, in Fig.\ \ref{fig:dens_mag_xy} we show
cross sections through the centre of runs B3-MH and B4-MH along the
$(x,y)$ plane, showing the density field and the component of the
magnetic field on this plane. We observe that in run B3-MH the field
lines are not strongly bent, except at the sites of local collapse. This
suggests that the M2FRs we measure by the projection method should not
exceedingly overestimate the actual flux-tube value.

Nevertheless, a more definite way to estimate the amount of
overestimation of the M2FR incurred in by the projection method is to
use a different estimator. One such estimator is what we call the
``local method'', which consists in measuring the M2FR for individual
cells in the simulations, using the total magnetic field strength.
Specifically, we measure the gas mass $M$ and read off the total
magnetic field strength $B$ in a cell, in order to calculate the M2FR as
\beq
\mu \approx \frac{M}{B dx^2},
\label{eq:mu_local}
\eeq
where $dx$ is the cell's side length. This estimator is gives a lower
bound to the M2FR in a magnetic flux tube, since it only counts the mass
in a single cell within that tube. This method has no observational
analogue but, by using the two methods, we expect to bracket the true
distribution of values of the M2FR, at least in a statistical sense.

\section{Results} \label{sec:results}

\subsection{Global evolution and star formation} \label{sec:global}

\begin{figure*}
\includegraphics[height=0.31\hsize]{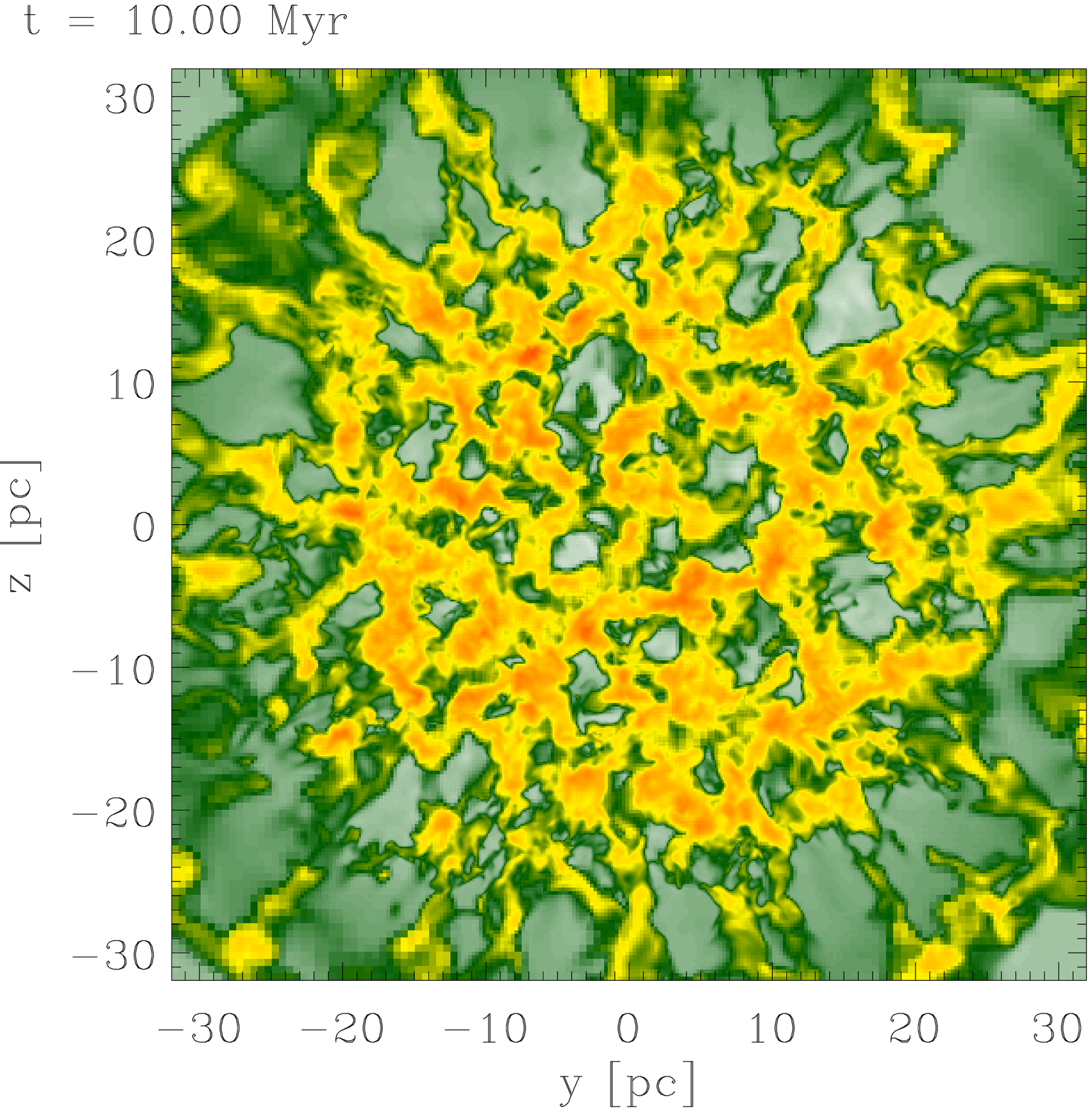}
\includegraphics[height=0.31\hsize]{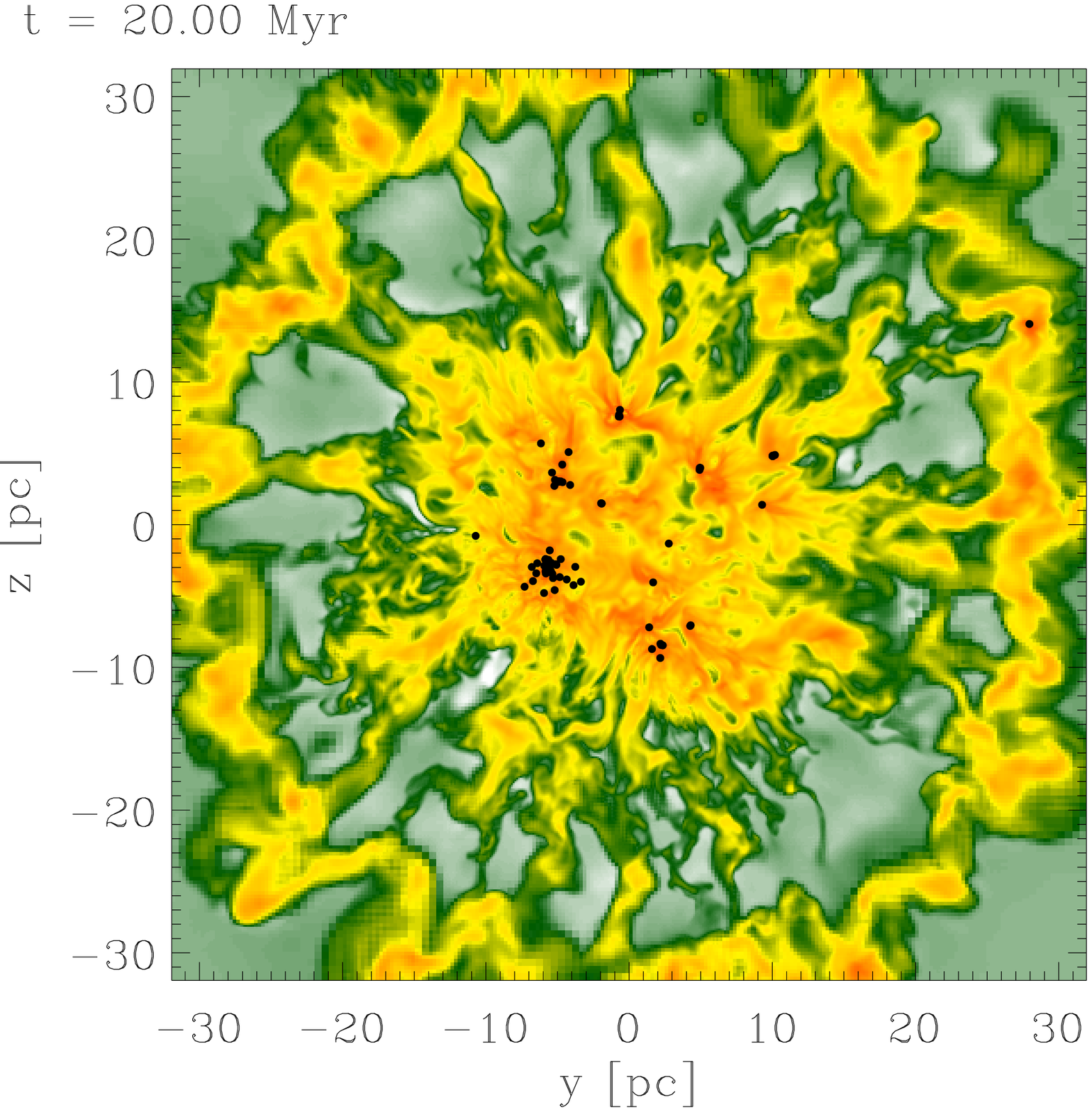}
\includegraphics[height=0.31\hsize]{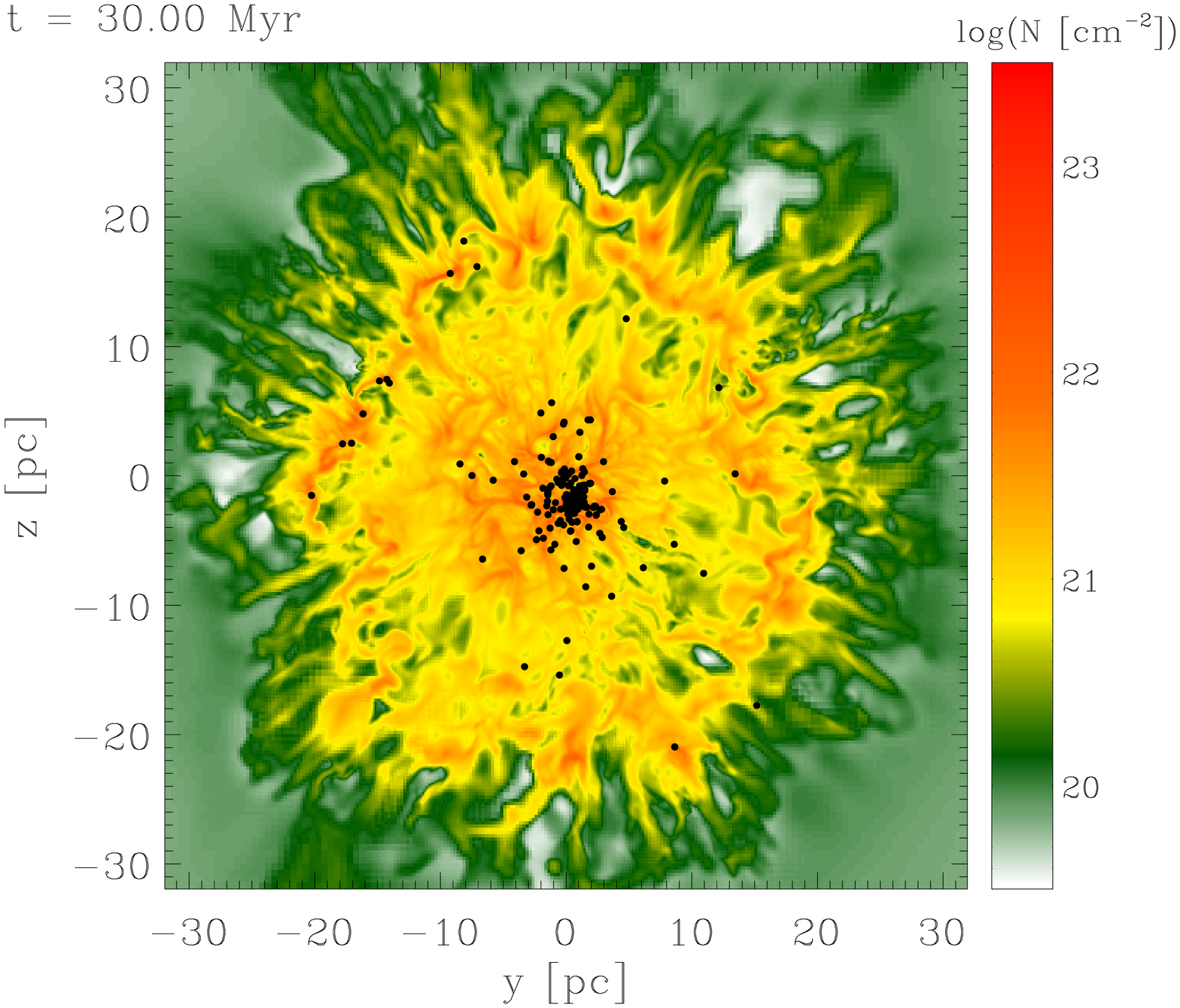}
\caption{Snapshots of run B2-AD showing the column density integrated
along the central 20 parsecs of the simulation along the $x$-direction
(perpendicular to the colliding inflows), at times $t=10$ Myr ({\it
left panel}), $t =20$ Myr ({\it middle panel}), and $t=30$ Myr ({\it
right panel}). }
\label{fig:run11_evol}
\end{figure*}

We first direct our attention to the global evolution of the clouds.
The supercritical run B2-AD evolves very similarly to the non-magnetic
runs presented by \citet{VS_etal07} and \citet{VS_etal10} and the
strongly supercritical runs presented by \citet{Henneb_etal08} and
\citet{Banerjee_etal09}. The cloud
starts out as a thin cylindrical sheet that fragments and thickens as
time increases, until it becomes gravitationally unstable and begins a
global radial contraction at $t \sim 9$ Myr. Shortly after that ($t \sim
12$ Myr) star formation begins in the densest fragments (``clumps''),
while the fragments continue to fall towards the global centre of mass,
and by $t \sim 24$ Myr a dense cloud of radius $\sim 10$ pc has formed
there, which does not appear to contract further.  This lack of
contraction, however, is only apparent, because in fact gas is being
consumed within the cloud by SF, and gas from the outside continues to
fall onto the cloud. Figure \ref{fig:run11_evol} shows this run at $t =
10$, 20 and 30 Myr, illustrating its evolution.

\begin{figure*}
\includegraphics[height=0.31\hsize]{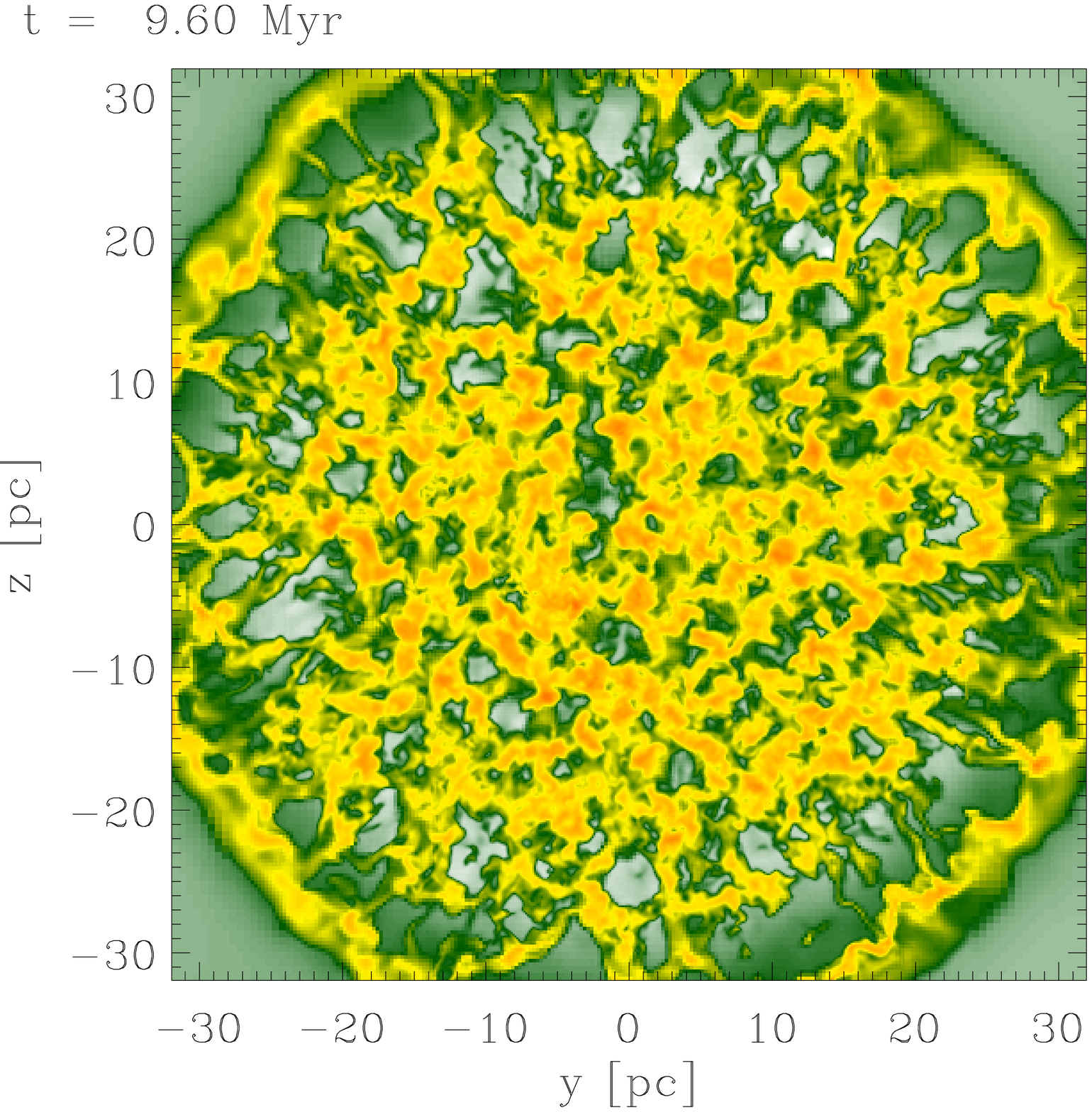}
\includegraphics[height=0.31\hsize]{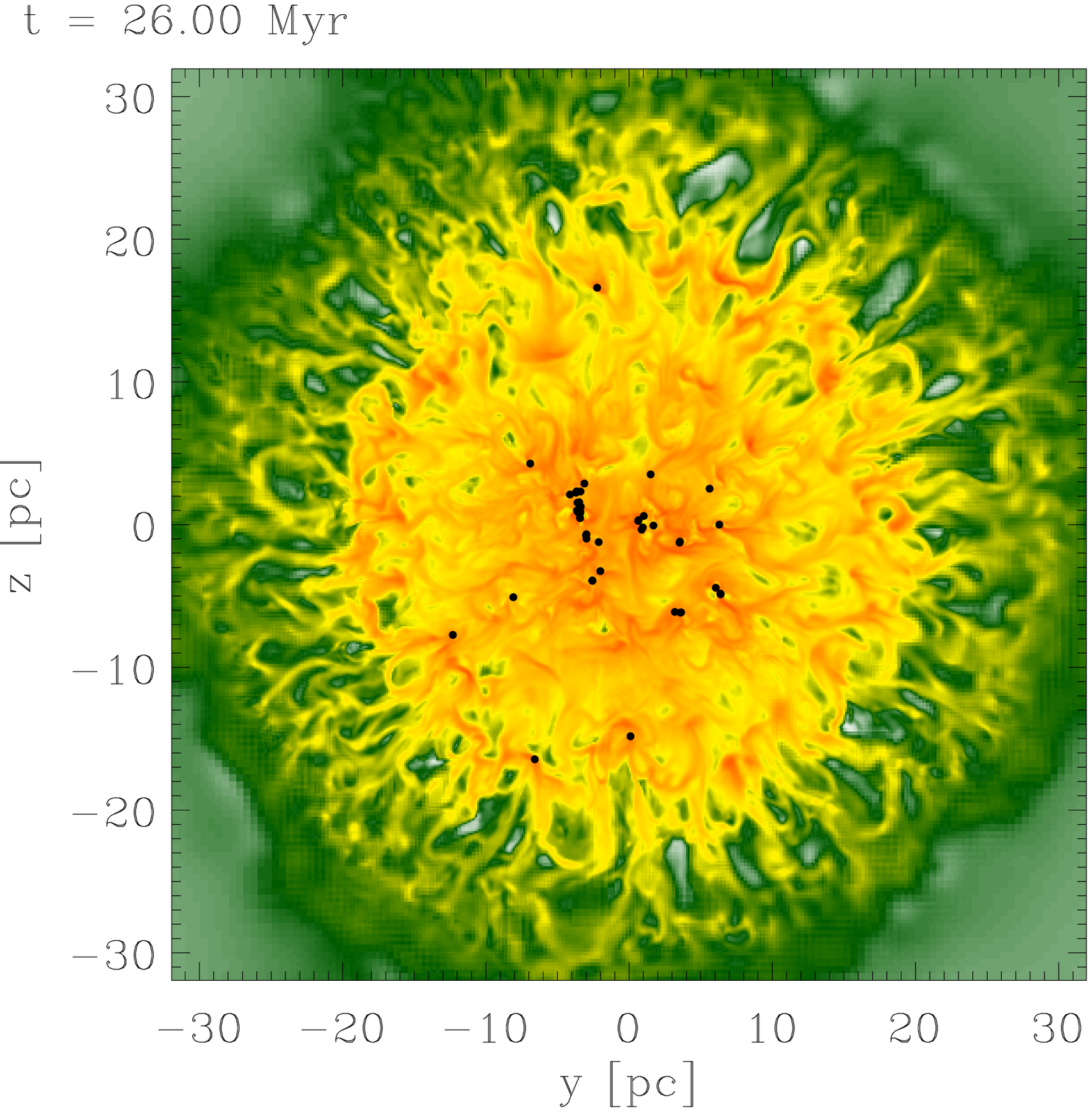}
\includegraphics[height=0.31\hsize]{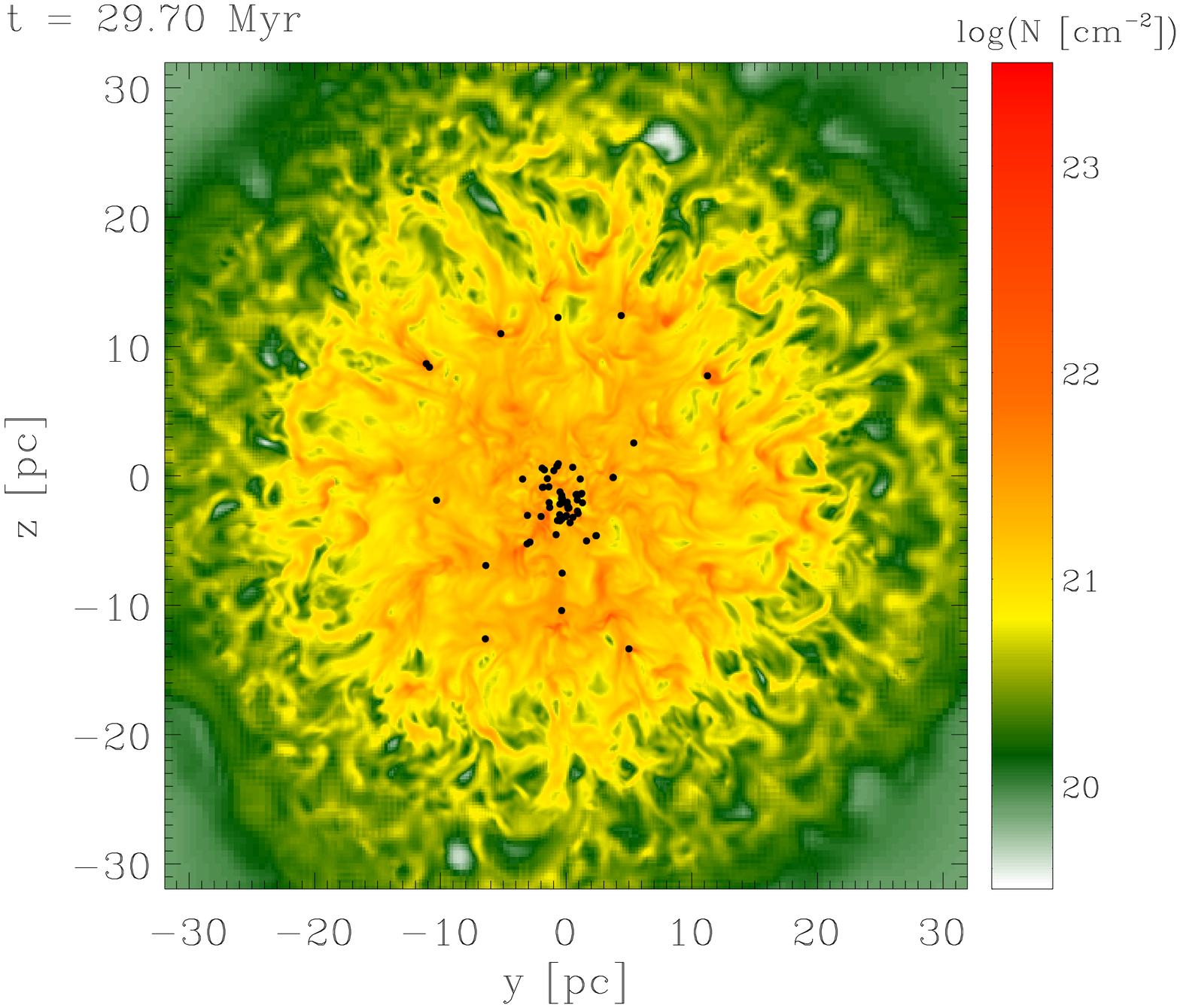}
\caption{Snapshots of run B3-AD showing the column density integrated
along the central 20 parsecs of the simulation along the $x$-direction
(perpendicular to the colliding inflows), at times $t=9.6$ Myr ({\it
left panel}), $t =26$ Myr ({\it middle panel}), and $t=29.7$ Myr ({\it
right panel}). The simulation contracts until $t=26$ Myr, after which it
begins to rebound.}
\label{fig:run08_evol}
\end{figure*}

\begin{figure*}
\includegraphics[width=0.45\hsize]{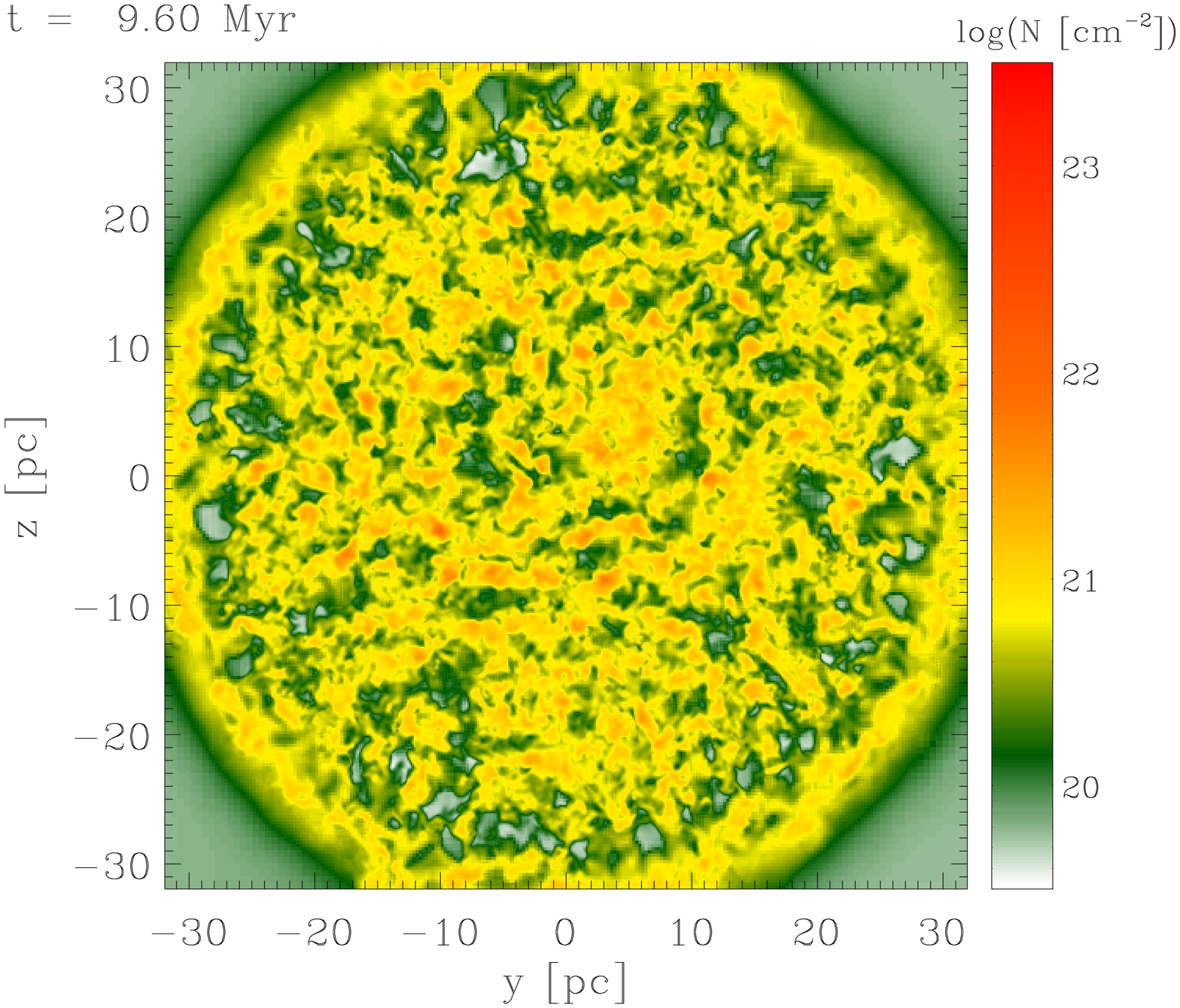}
\includegraphics[width=0.45\hsize]{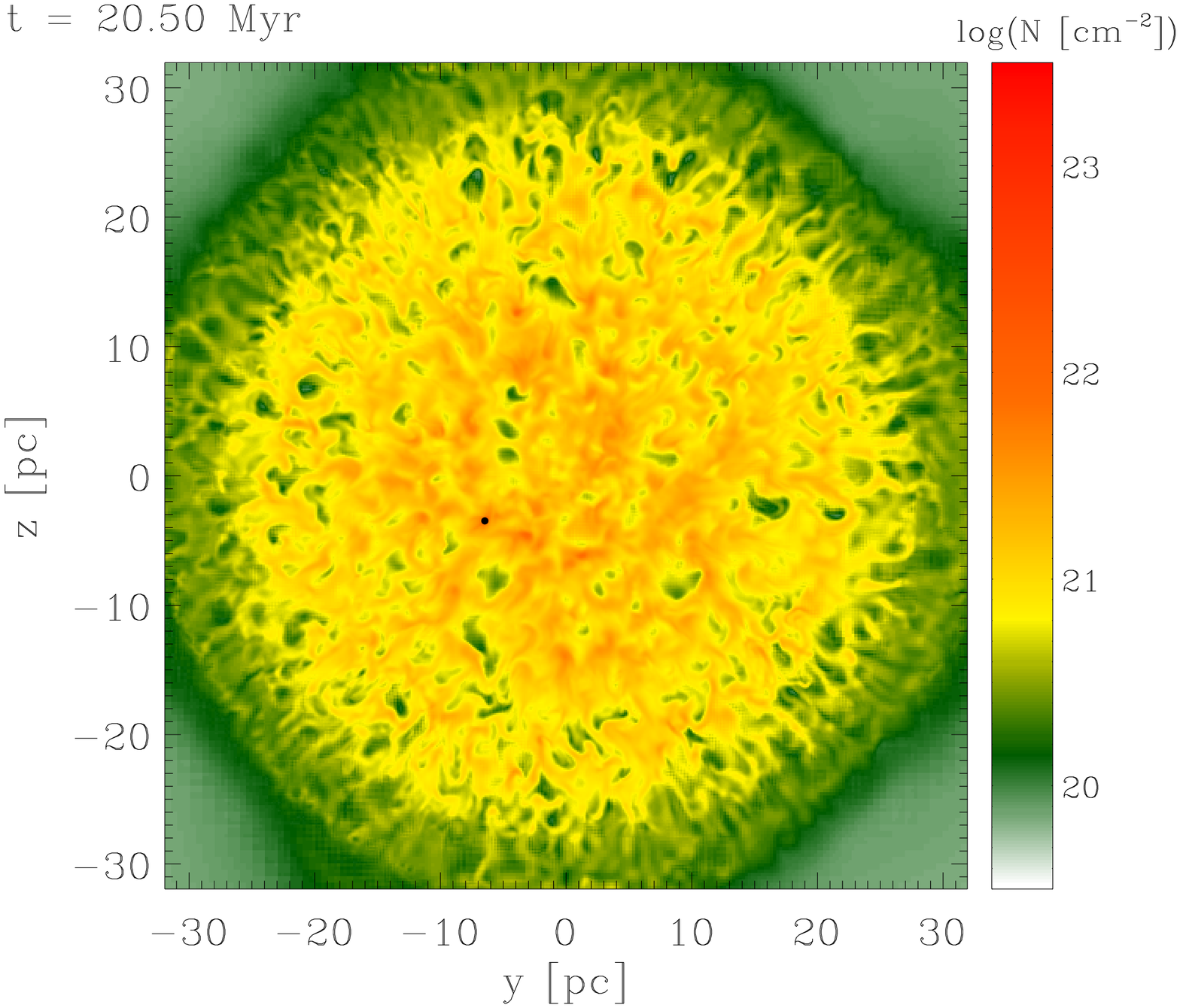}
\includegraphics[width=0.45\hsize]{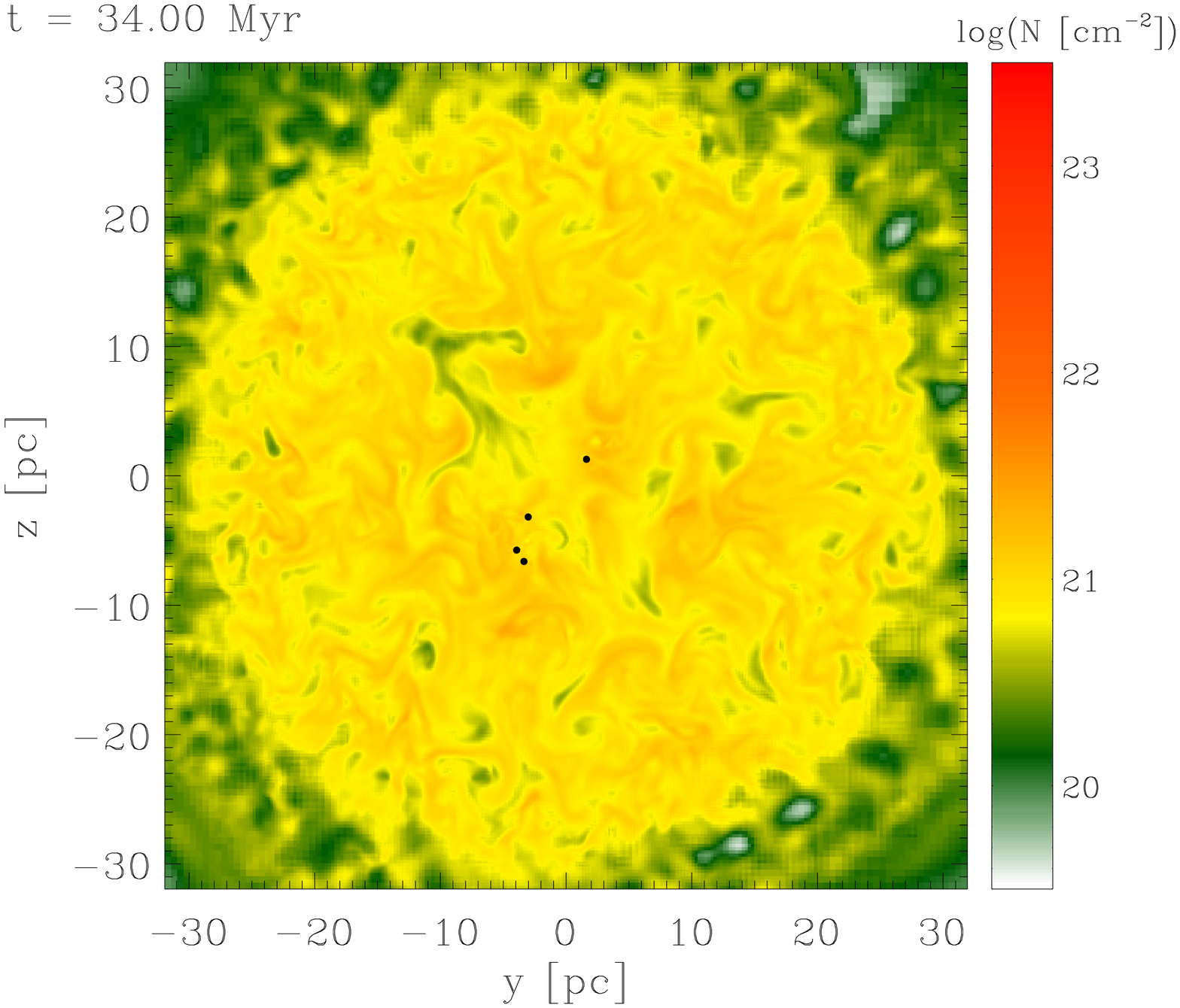}
\includegraphics[width=0.45\hsize]{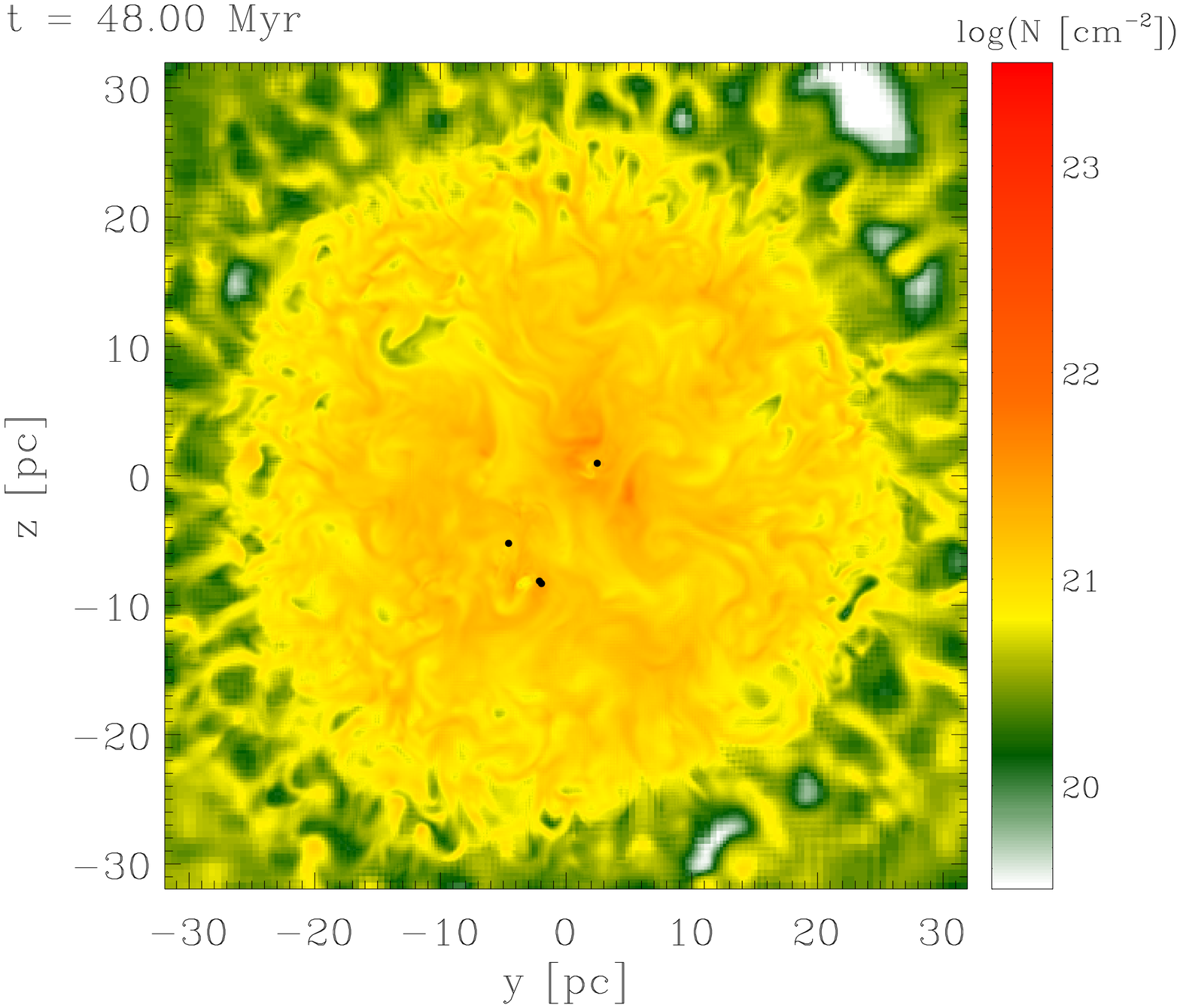}
\caption{Snapshots of run B4-AD showing the column density integrated
along the central 20 parsecs of the simulation along the $x$-direction
(perpendicular to the colliding inflows), at times
$t=9.6$ Myr ({\it top left panel}) $t=20.5$ Myr ({\it top right panel}),
$t= 34$ Myr ({\it bottom left panel}), and $t = 48$ Myr ({\it bottom
right panel}). The simulation is seen to be undergoing global
oscillations, with star formation (indicated by the dots) scarcely
occurring only during the times of maximum compression.}
\label{fig:run10_evol}
\end{figure*}

\begin{figure*}
\includegraphics[width=0.45\hsize]{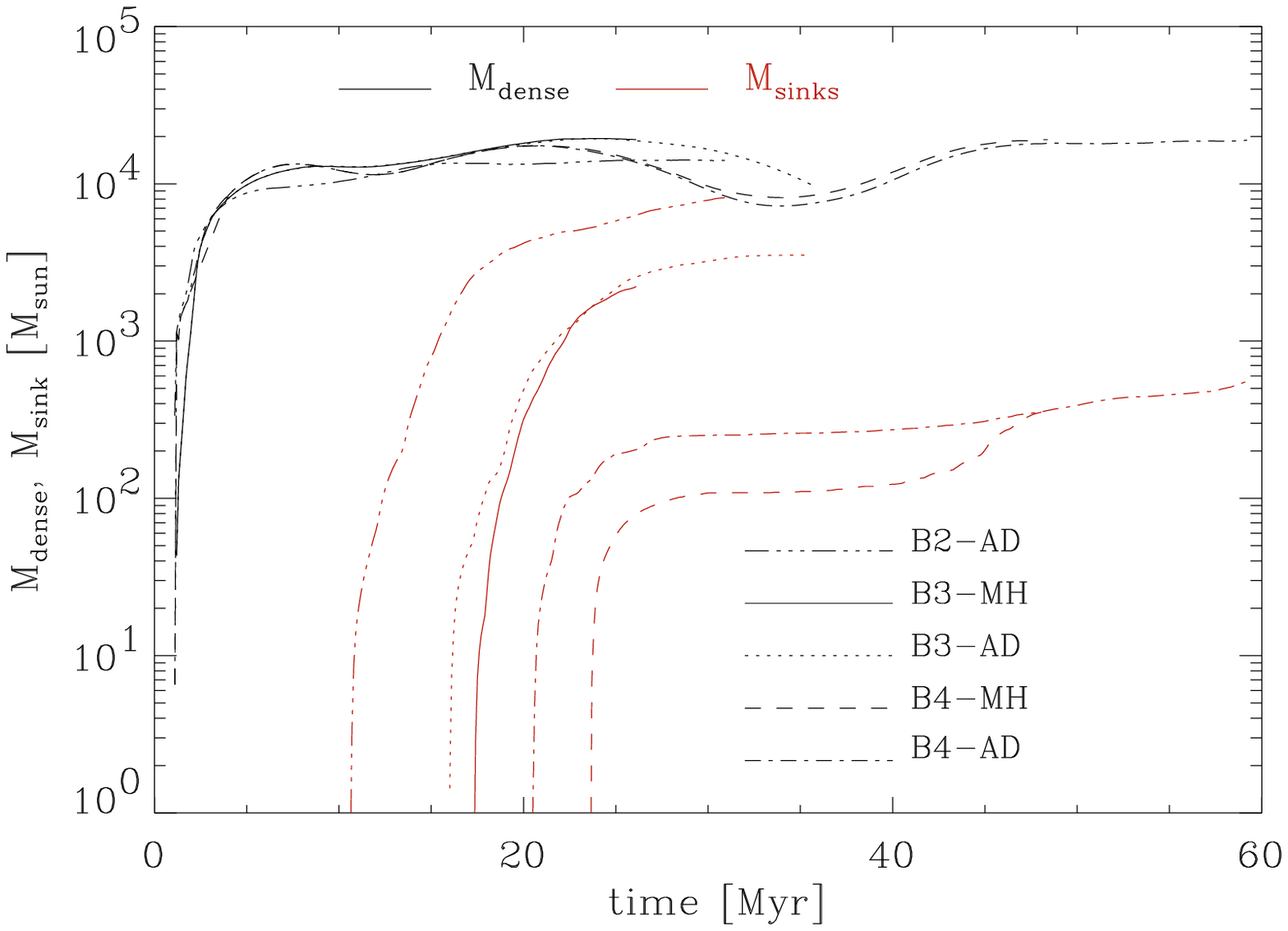}
\includegraphics[width=0.45\hsize]{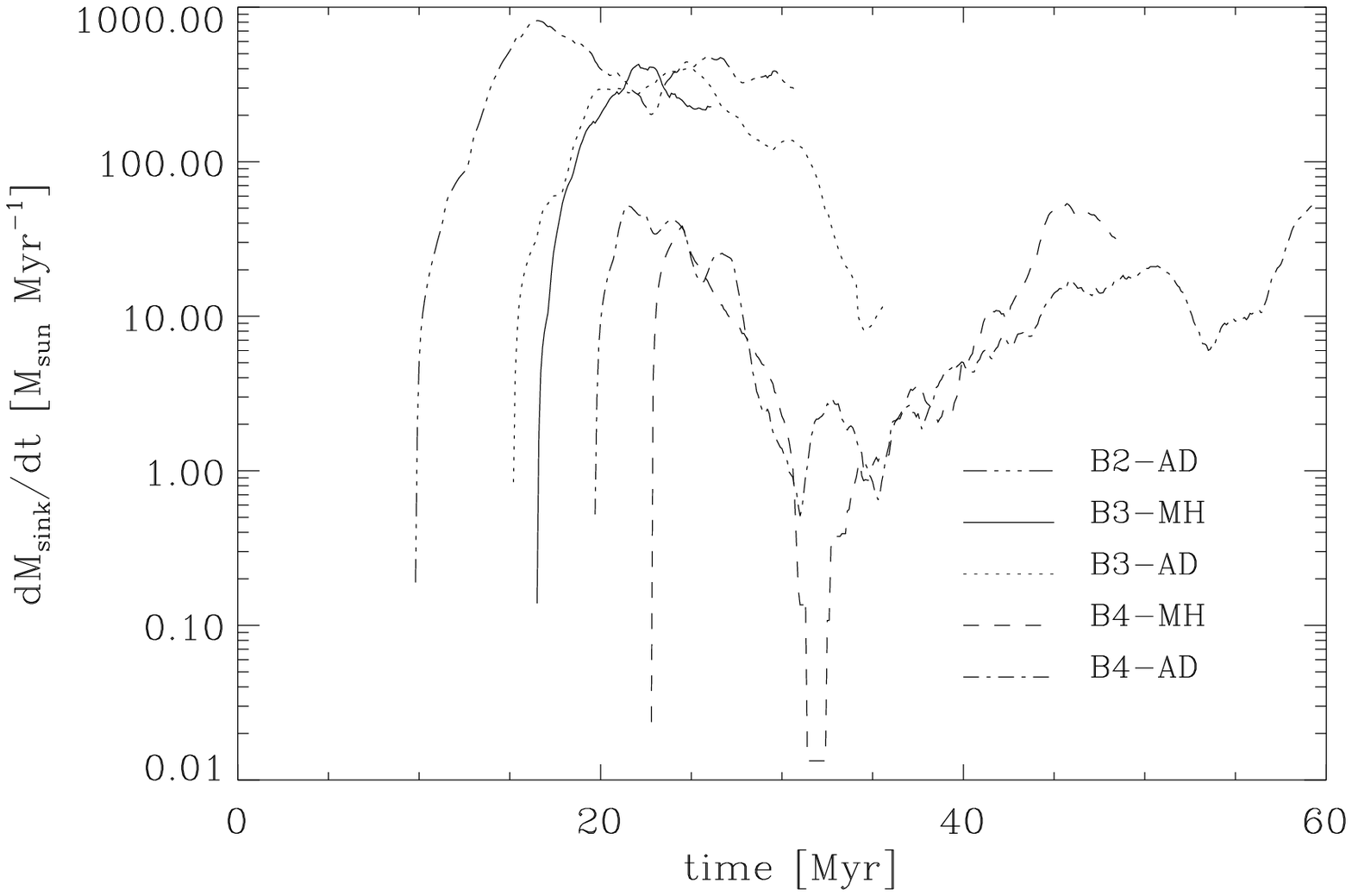}
\includegraphics[width=0.45\hsize]{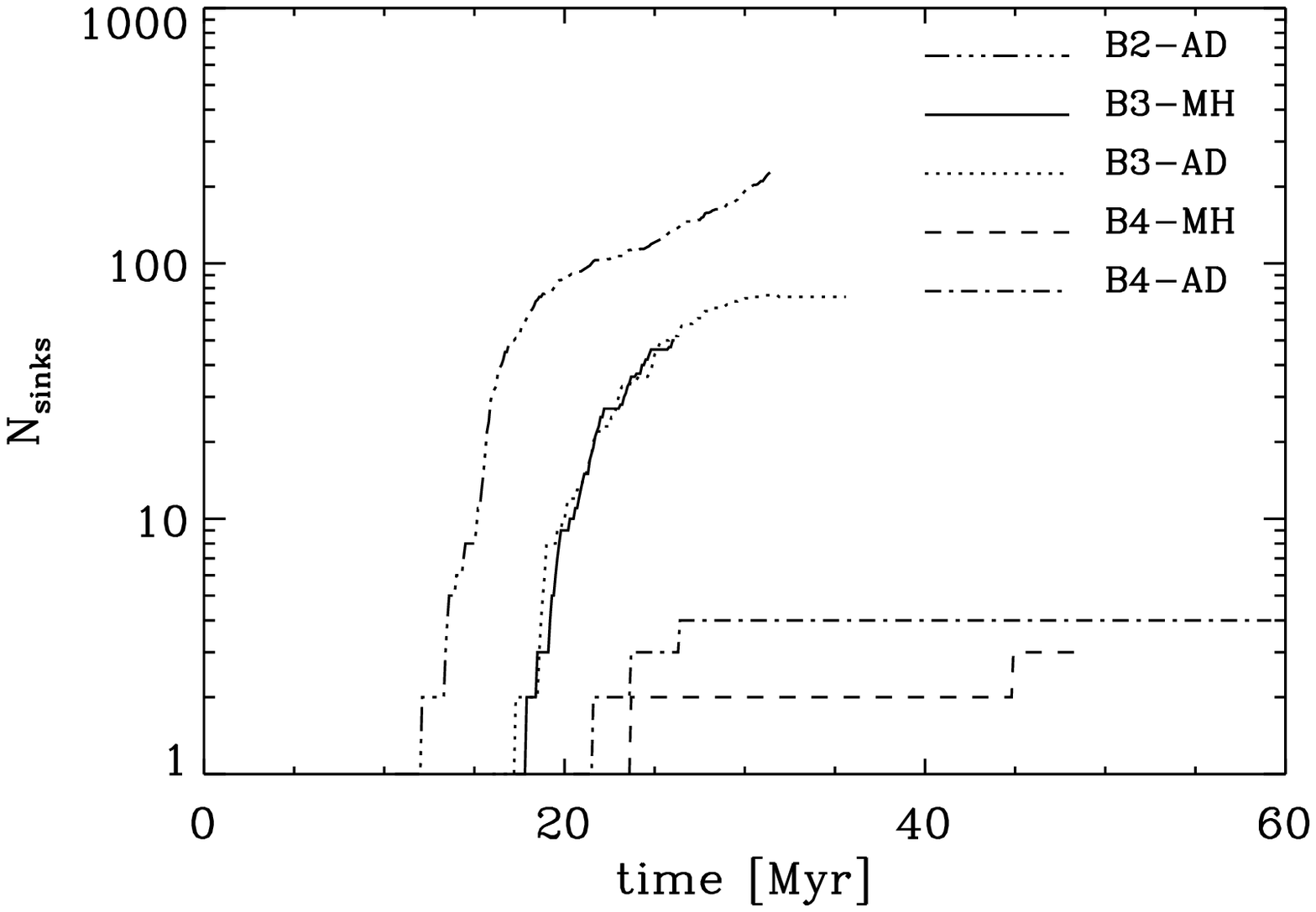}
\includegraphics[width=0.45\hsize]{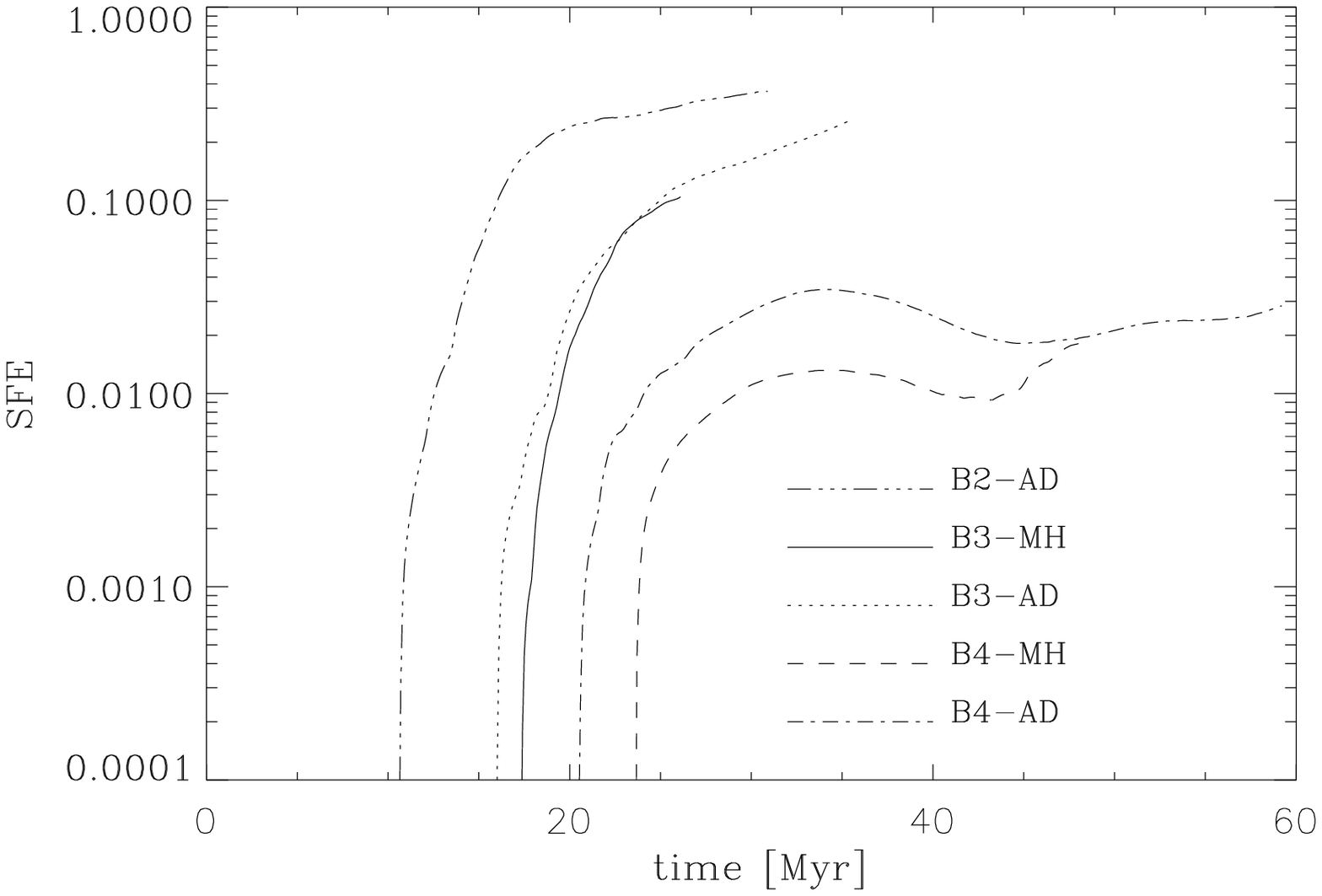}
\caption{Evolution of the total dense gas mass ($n > 100 \pcc$) and
total sink mass ({\it top left panel}), the time derivative of the total
sink mass $\Msinkdot$ (approximately giving the SFR; {\it top right
panel}), the total number of sink particles ({\it bottom left
panel}), and the SFE, defined by eq.\ (\ref{eq:SFE}) ({\it bottom right
panel}) in all simulations. The density range for the dense gas includes
gas that would classify as ``atomic'' as well as ``molecular''. The
graphs of $\Msinkdot$ have been smoothed by averaging ovser 16 neighboring
data points at each plotted value.}
\label{fig:sink_gas_SFE_evol}
\end{figure*}

On the other hand, the subcritical runs B3 and B4 undergo a period of
initial contraction followed by a {\it rebound}, eventually settling
into an oscillatory regime, which consists of alternating periods of
contraction and expansion around the magnetostatic equilibrium state, as
previously observed by \citet{LN04}. These oscillations are best seen in
animations of the simulations (not shown), but they can also be observed
in Figs.\ \ref{fig:run08_evol} and \ref{fig:run10_evol}, which show
snapshots of the density field of runs AD-B3 and AD-B4 at various times,
respectively. In Fig.\ \ref{fig:run08_evol}, it can be seen that the
central density is larger at the intermediate time shown in the {\it
middle} panel than at the final time shown at the {\it right} panel. A
similar behavior is seen in Fig.\ \ref{fig:run10_evol}, where the
central density is seen to be larger at times $t=20.5$ and $t=48$ Myr
than at $t=34$ Myr. The B3 runs, having a weaker mean field,
contract for a longer time (up to $t \sim 25$ Myr) and reach a smaller
size (radius $R \sim 20$ pc) than the B4 runs (maximum contraction at $t
\sim 20$ Myr, with radius $R \sim 25$ pc). The B4 runs, which were
followed to longer times, clearly exhibit the oscillatory regime, with a
period of $\sim 30$ Myr. The oscillations can also be seen in Fig.\
\ref{fig:sink_gas_SFE_evol}, which we now discuss.

Figure \ref{fig:sink_gas_SFE_evol} shows, for all the runs, the
evolution of the total dense gas mass and total sink mass ({\it top left
panel}), the time derivative of the total sink mass $\Msinkdot$ ({\it
top right panel}), the total number of sinks ({\it bottom left panel}),
and the SFE ({\it bottom right panel}), defined as 
\beq
\hbox{SFE}(t) = \frac{\Msinks(t)}{\Mdense(t) + \Msinks(t)},
\label{eq:SFE}
\eeq
where $\Mdense$ is the mass of the gas with density $n > 100 \pcc$, and
$\Msinks$ is the total mass in sink particles. We take  $\Msinkdot$ as a
proxy for the SFR.

The evolution of the runs is seen to depend sensitively on the magnetic
field strength. For $t \ga 7$ Myr, the dense gas mass oscillates by
factors of 2--4 for the subcritical runs B3 and B4, in both the MHD and
the AD cases, evidencing again the oscillatory regime in which these
runs engage. The maxima of the mass in the subcritical runs coincide
with the times of maximum compression. This may be partially an artifact
of the threshold density we have chosen for defining the cold
gas. Because the global magnetic support for the cloud prevents it from
contracting much, the gravitational binding of the cloud is generally
weak. This in turn means that the during periods of maximum expansion, a
significant fraction of the cloud, although still in the cold phase, may
be below the density threshold of $100 \pcc$ we have used for defining
the cloud. Nevertheless, if the mean density of the cloud varies, it is
likely that the molecular fraction should actually vary as well, since
molecular gas may be dissociated if the cloud's column density decreases
sufficiently \citep{Glover_etal10}. Thus, during periods of expansion,
the cloud may contain a 
lower molecular fraction. In addition, significant amounts of gas may be
in a transient state between the warm and cold phases of the ISM
\citep[see the reviews by][ and references therein]{VS_etal03, HMV09, VS09},
being thus lost from the cold, dense phase.

The supercritical run B2, on the other hand, does not exhibit such
strong oscillations in its dense gas mass content. Instead, the dense
gas mass increases rapidly at first. This is because this run does not
engage in any radial oscillations, but simply proceeds directly to
collapse. Interestingly, however, the dense gas mass later becomes
roughly stationary, although this stabilization is not due to the cloud
being in any sort of equilibrium, but rather to the fact that it is
forming stars at roughly the same rate it accretes mass from the
WNM. Indeed, the {\it red lines} in the {\it top left panel}, as well as
the {\it top right panel} of Fig.\ \ref{fig:sink_gas_SFE_evol} show that
the total sink mass in the simulation increases at a steady pace, of
roughly $400 \Msun$ Myr$^{-1}$ during the time interval $20 < t < 30$ Myr.

All runs, including those that are subcritical, form ``stars'' (i.e.,
sink particles), since, as discussed in Sec.\ \ref{sec:resolution},
numerical diffusion acts in a similar manner to AD. Nevertheless, 
when AD is included the total sink number and mass increase, the dense
gas mass decreases, these effects being relatively more noticeable for
the strong-field runs B4. In general, as can be seen from the {\it
bottom left} panel of Fig.\ \ref{fig:sink_gas_SFE_evol}, the total
number of sinks in the B4 and B3 cases is larger only by a few sinks
when AD is included.  This reinforces our conclusion from Sec.\
\ref{sec:resolution} that numerical diffusion has an effect of
comparable strength to that of AD in our simulations, since AD is able
to induce the collapse of a few clumps in addition to those that
collapse because of numerical diffusion.  In the case of the B3 runs,
the relative effect of AD is smaller because the same difference of a
few extra sink particles is a smaller fraction of the total number of
sinks produced.

Indeed, run B3-AD forms stars at a much higher rate than run
B4-AD, even though both are subcritical. The time derivative of the sink
mass, $\Msinkdot$, in run B3-AD is roughly one order of magnitude larger
than that of run B4-AD, as seen in the {\it top right panel} of Fig.\
\ref{fig:sink_gas_SFE_evol}. In fact, the formation of sinks completely
stops in run B4-AD after $t \approx 27$, as can be seen in the {\it
bottom left panel} of this figure. The very slight increase in the sink
mass observed during this period ({\it top left panel, dash-dotted red
line}) is due to accretion onto the existing sink particles, rather than
to the formation of new particles. Run B2-AD, on the other hand, is seen
to have a larger SFR than run B3-AD, although only by factors of a
few. Also, it can be seen from all panels of Fig.\
\ref{fig:sink_gas_SFE_evol} that the onset of SF is delayed as $B_0$ is
increases, but that the presence of AD shortens this delay.  It is
important to note that all SF in the runs occurs {\it after} the cloud
has been assembled and its M2FR has reached a nearly stationary value
($t > 10$ Myr).

To conclude this section, the {\it bottom right panel} of Fig.\
\ref{fig:sink_gas_SFE_evol} shows the evolution of the SFE, defined by
eq.\ (\ref{eq:SFE}), in all simulations. Again, a continuous trend of
increasing SFE with decreasing $B_0$ is observed throughout our set of
simulations. The supercritical run B2-AD reaches an SFE of $\sim 35$\%
at $t= 30$ Myr. For comparison, at this time, run B3-AD has reached an
SFE of $\sim 25$\%, while run B4-AD has reached only SFE $\sim
3$\%. Although at face value these numbers would suggest that run B4-AD
compares best to the observed SFEs of GMCs \citep{Myers_etal86,
Evans_etal09}, this conclusion may be premature, since the additional
effect of stellar feedback in reducing the SFE
\citep[e.g.,][]{VS_etal10} is not taken into account in the present
simulations. We conclude that the SFR and the SFE can depend sensitively
on the mean field strength, even for globally subcritical cases, and
that the marginally subcritical run has an SFE comparable to that of the
supercritical case.

\subsection{Spatial and probability distribution of the mass-to-flux ratio}
\label{sec:M2FR_distr}

\begin{figure*}
\includegraphics[height=0.31\hsize]{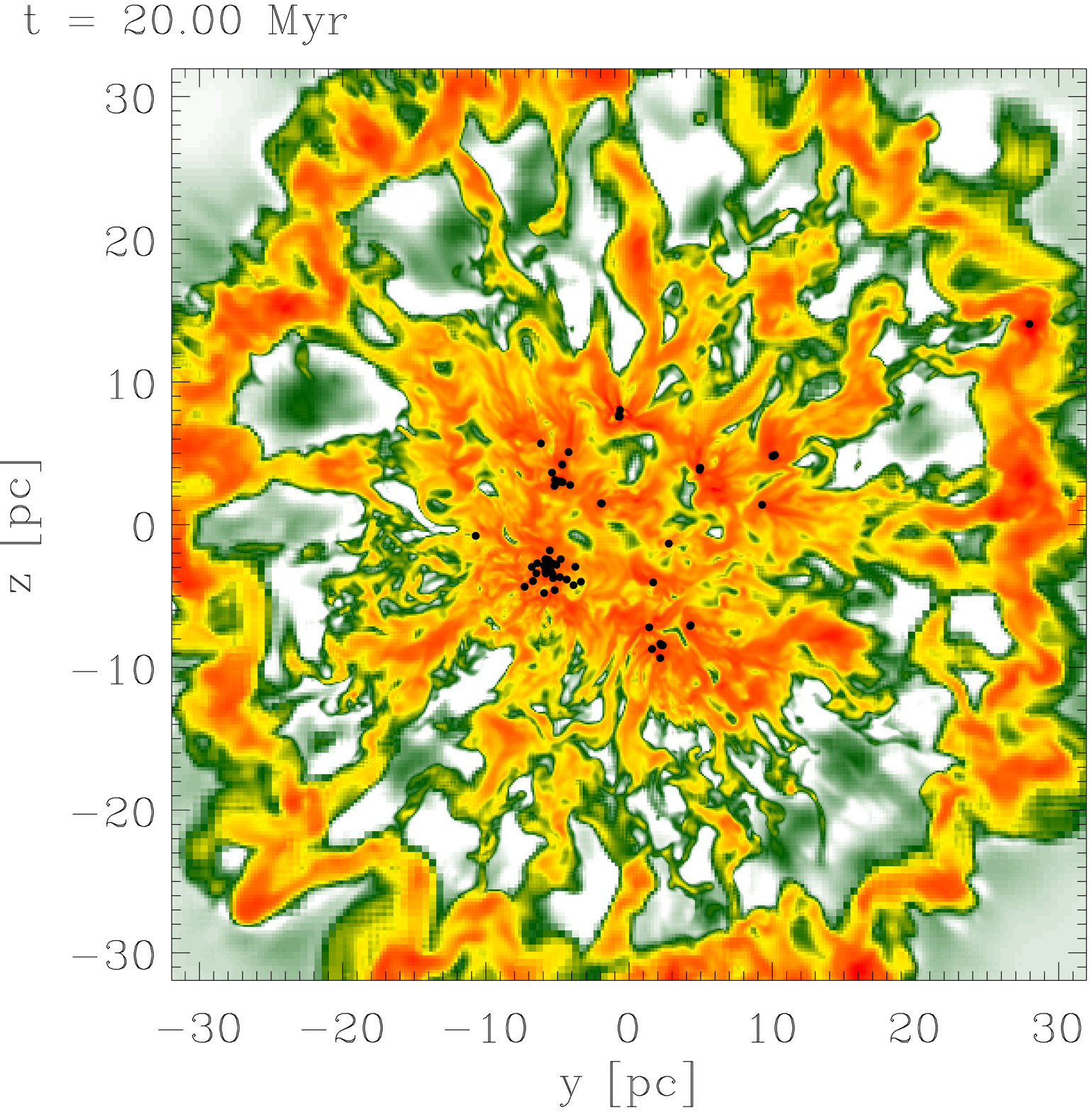}
\includegraphics[height=0.31\hsize]{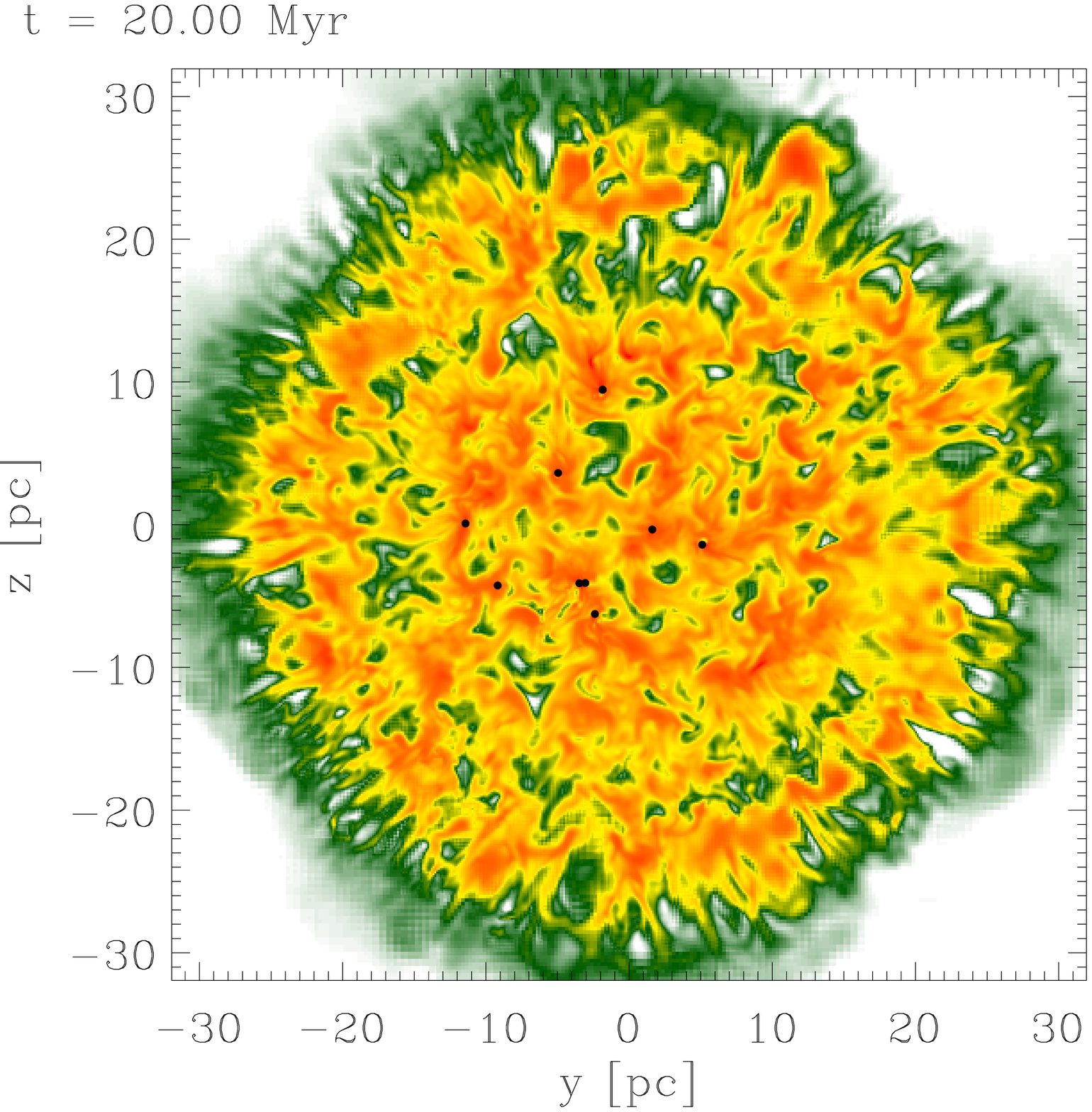}
\includegraphics[height=0.31\hsize]{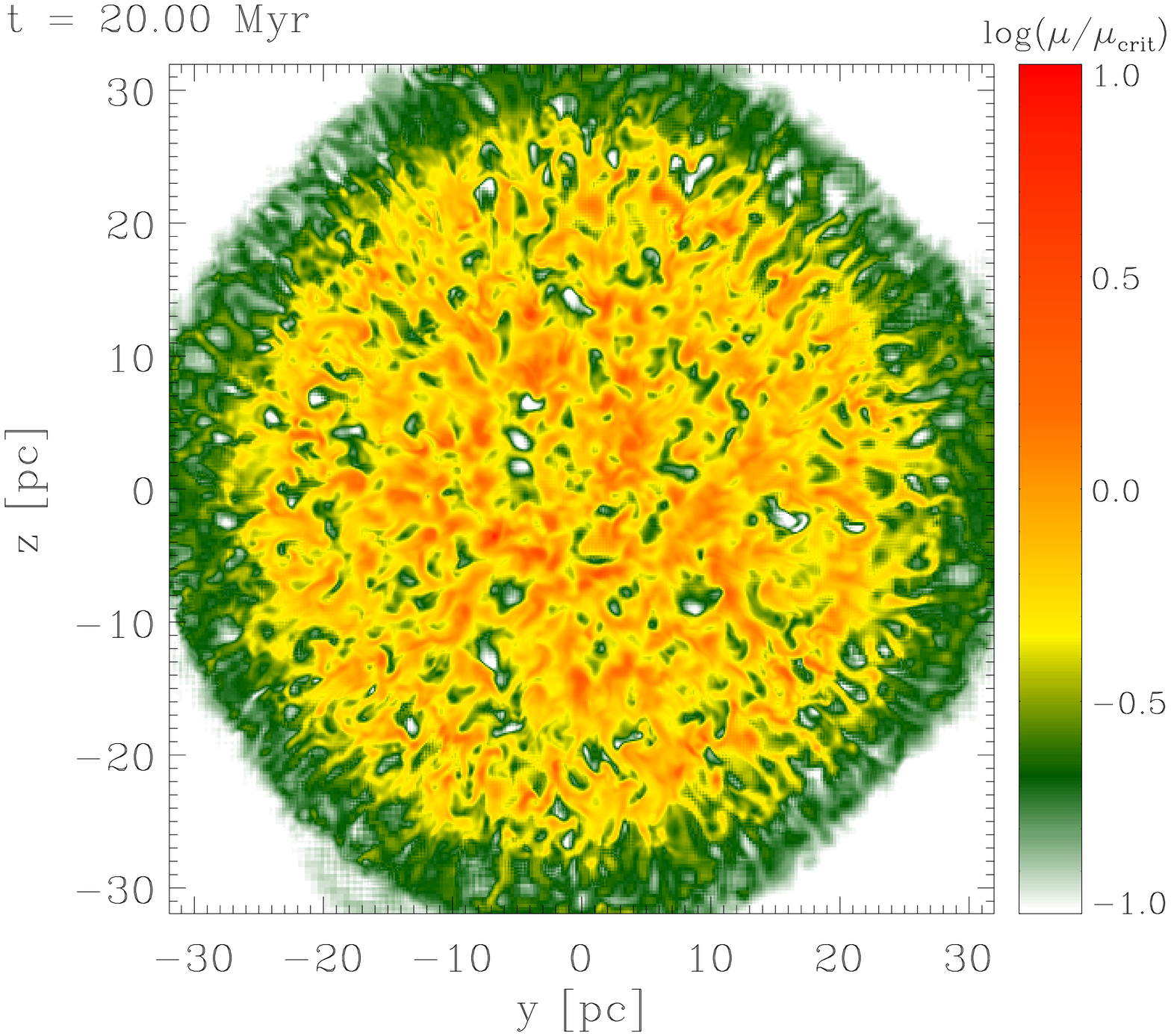}
\hglue 0.31\hsize
\includegraphics[height=0.31\hsize]{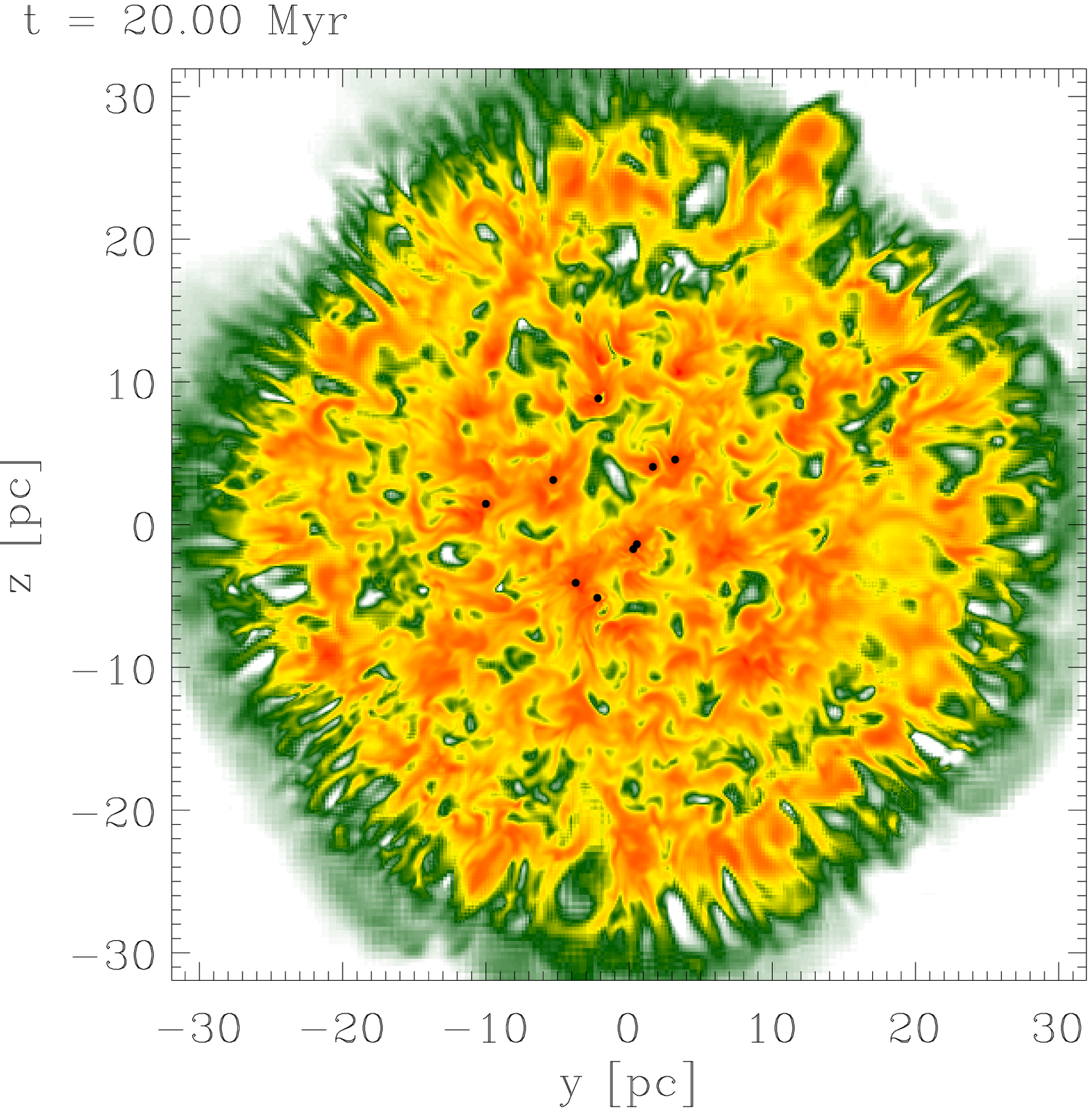}
\includegraphics[height=0.31\hsize]{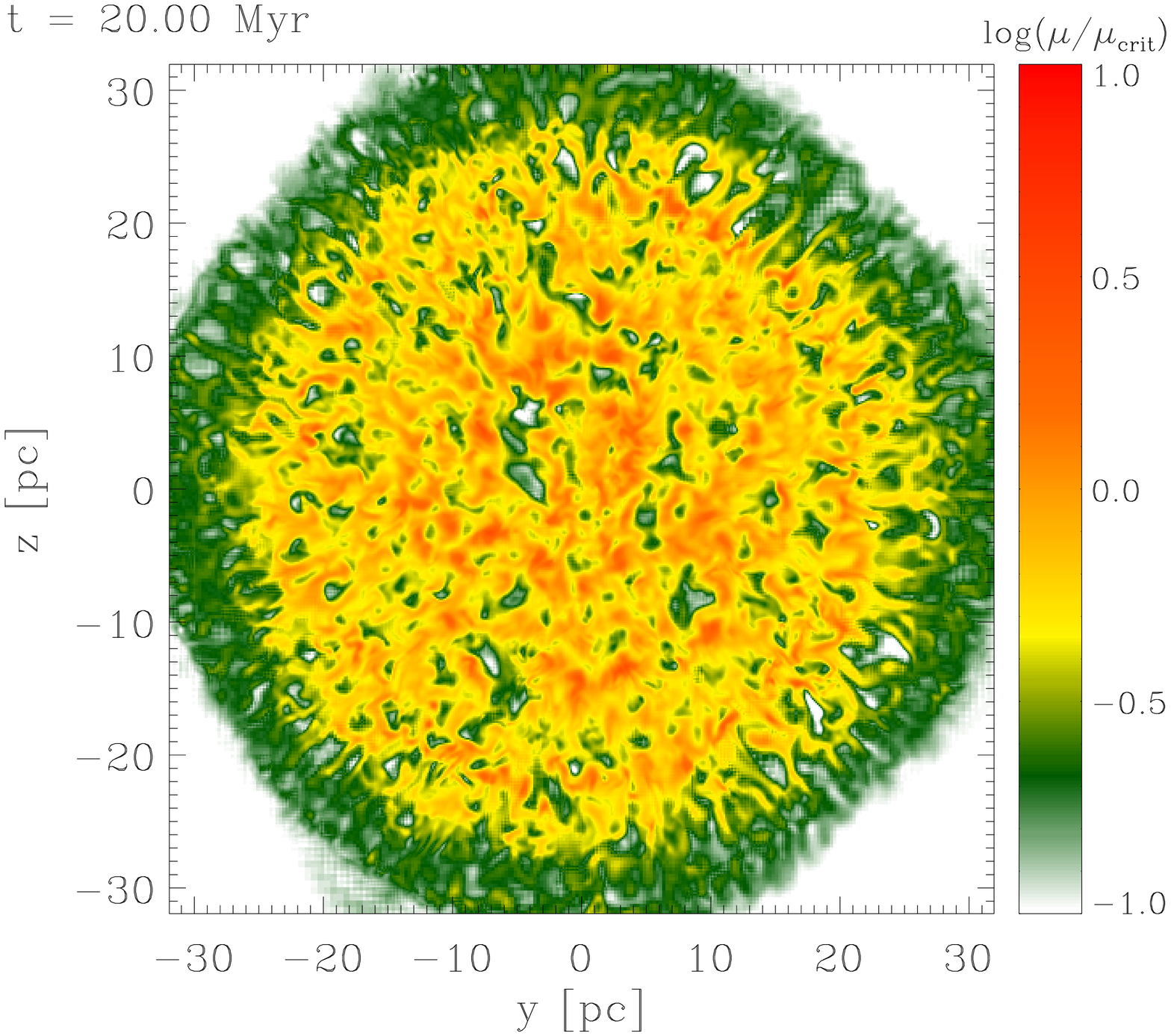}
\caption{Snapshots of the mass-to-flux ratio $\mu$, normalized to the
critical value, for runs B2-AD ({\it top left panel}), B3-AD ({\it top middle
panel}), B4-AD ({\it top right panel}), B3-MH ({\it bottom middle
panel}) and B4-MH ({\it bottom right panel}) at time $t=20$ Myr. The M2FR is
calculated as $N/B_{\rm LOS}$, as indicated by eq.\
(\ref{eq:proj_mu}), integrated over a 20-pc path centered at the
midplane of the simulation along the direction of the inflows. This path
completely encloses the cloud along the $x$-direction in all
simulations. Note that there exist vast numbers of supercritical
filaments intermixed with subcritical patches. The latter, however,
occupy most of the volume. The dots indicate the positions of sink
particles. Note also that the density structure is similar at large
scales but different in the small-scale detail between the cases with
and without AD.}
\label{fig:mu_images}
\end{figure*}

A key piece of information needed to understand the behavior of
simulations is the spatial distribution of the M2FR, as well as the
evolution of its global average value. In what follows, we discuss the
M2FR estimated using the projection method (cf.\ eq.\
[\ref{eq:proj_mu}]) along LOSs parallel to the $x$ axis (i.e.,
perpendicular to the plane of the cloud), taking $L$ as the path $-10 <
x < 10$ pc along the direction of the inflows. We consider the M2FR
normalized to the critical value given by eq.\ (\ref{eq:NN78}).  Given
the flattened geometry of our clouds, we consider that this is a more
realistic value of the critical M2FR than the other commonly encountered
critical value of $(6 \pi^2 G)^{-1/2} \approx 0.13 G^{-1/2}$, which
holds for spherical geometry \citep[e.g.,] []{Shu92}. Figure
\ref{fig:mu_images} shows snapshots of the normalized M2FR for runs
B2-AD, B3-AD and B4-AD in the {\it top row}, and of runs B3-MH and
B4-MH in the {\it bottom row}, all at $t= 20$ Myr. In all cases, the
spatial distribution of the M2FR is seen to fluctuate strongly.

\begin{figure*}
\includegraphics[width=0.45\hsize]{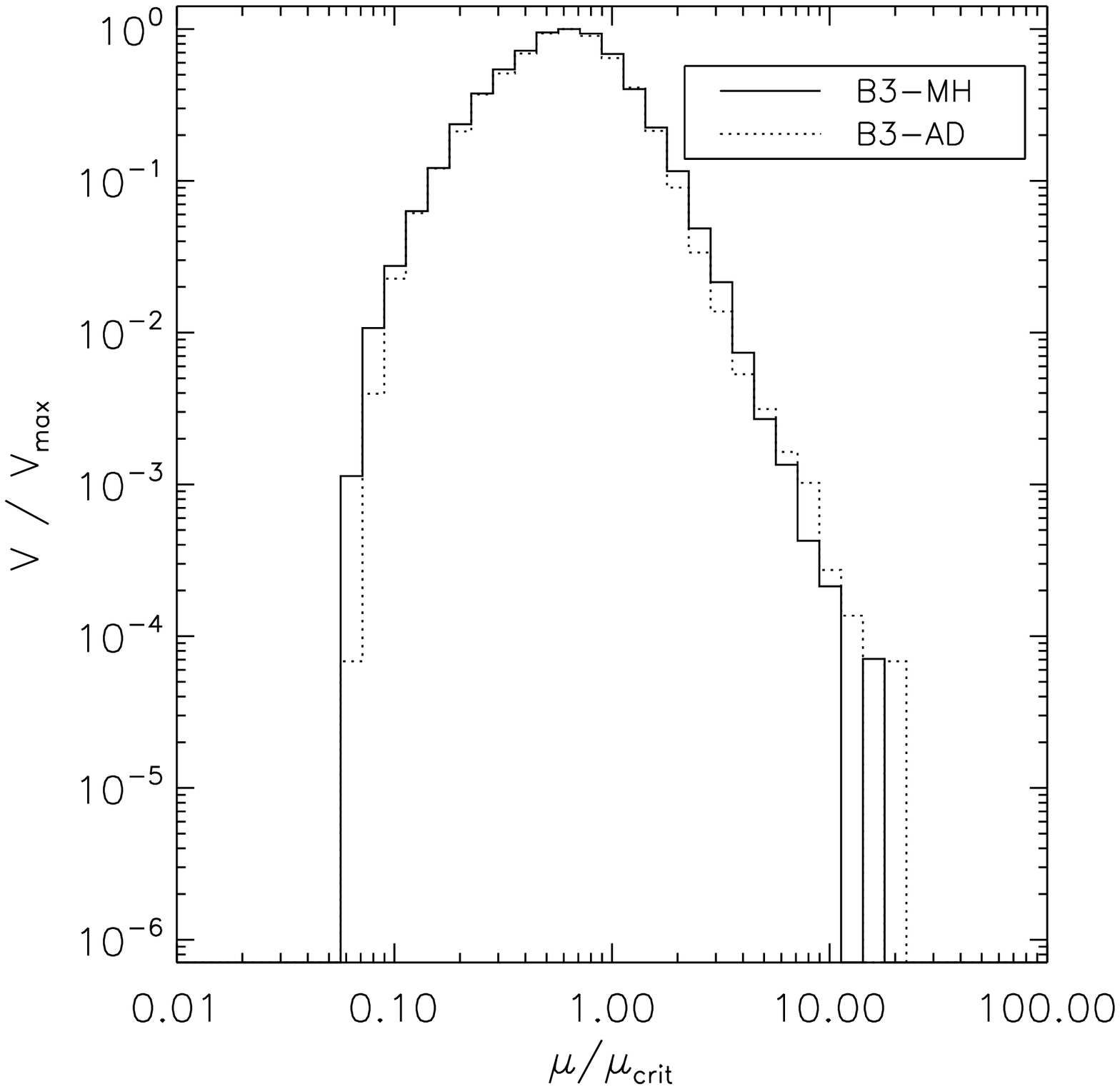}
\includegraphics[width=0.45\hsize]{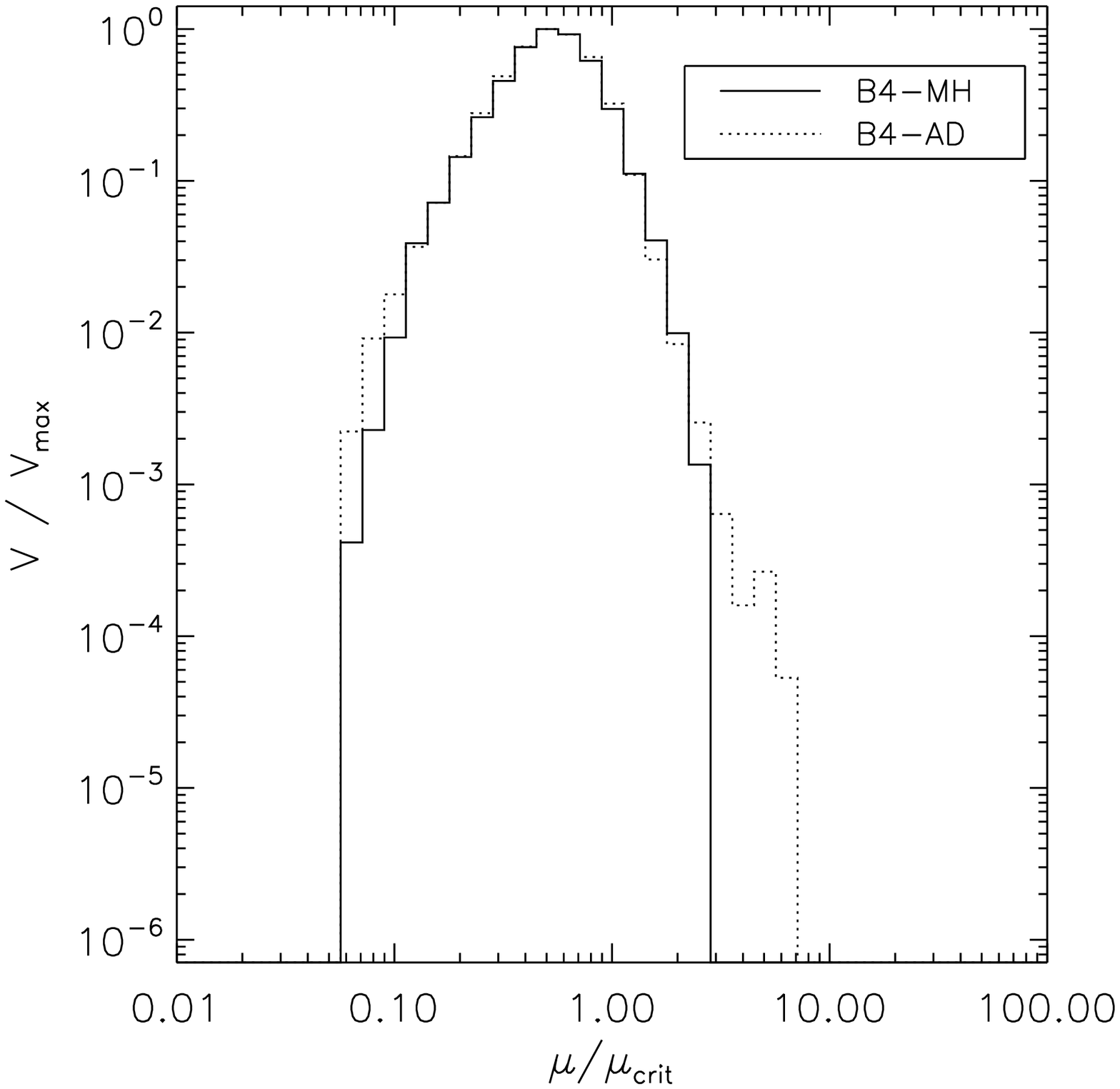}
\caption{{\it Left:} Comparison of the M2FR histograms at
$t=20$ Myr, for cases with ({\it dotted lines}) and without ({\it solid
lines}) AD, for runs B3 ({\it left panel}), and B4 ({\it right
panel}). The M2FR is estimated through the ``projection method'', as
indicated by eq.\ (\ref{eq:proj_mu}) over a circular region centered at
the point $(y,z) = (0,0)$, and of radius 20 pc. }
\label{fig:mu_histo_MH_vs_AD}
\end{figure*}

Comparing the MH and AD cases, it is interesting to note that the
spatial structure of the M2FR is similar in the large-scale
features, but differs in the shape and precise location of the fine,
small-scale ones. This is a manifestation of the system being chaotic,
so that the subtle variations in the magnetic forces at the densest
structures induced by AD are sufficient to change the details in the
topology of the gas. Presumably, for sufficiently long times the
differences in structure will reach even the largest scales. On the
other hand, simple visual inspection of the images is not enough to
discern any trend of systematically larger values of the M2FR in the
presence of AD. To quantify this, we show in Fig.\
\ref{fig:mu_histo_MH_vs_AD} the histograms of the M2FR. The histograms
are computed for all 
LOSs within a circular region centered at $(y,z) = (0,0)$, with a 20-pc
radius. From these, we see that the inclusion of AD causes the
production of a small excess of high-M2FR cells in comparison with the
non-AD cases, and that this effect is most noticeable in the
strong-field (B4) case, in which the maximum value of the M2FR is over a
factor of 2 larger than in the non-AD case. Instead, in the B3 case, the
excess is marginal, suggesting that when the system is very close to
being supercritical, numerical diffusion dominates over AD. This result
suggests that the relative importance of AD and numerical diffusion
depends on the mean field strength, an issue that deserves further
exploration, but which is out of the scope of the present paper.

\begin{figure*}
\includegraphics[width=0.45\hsize]{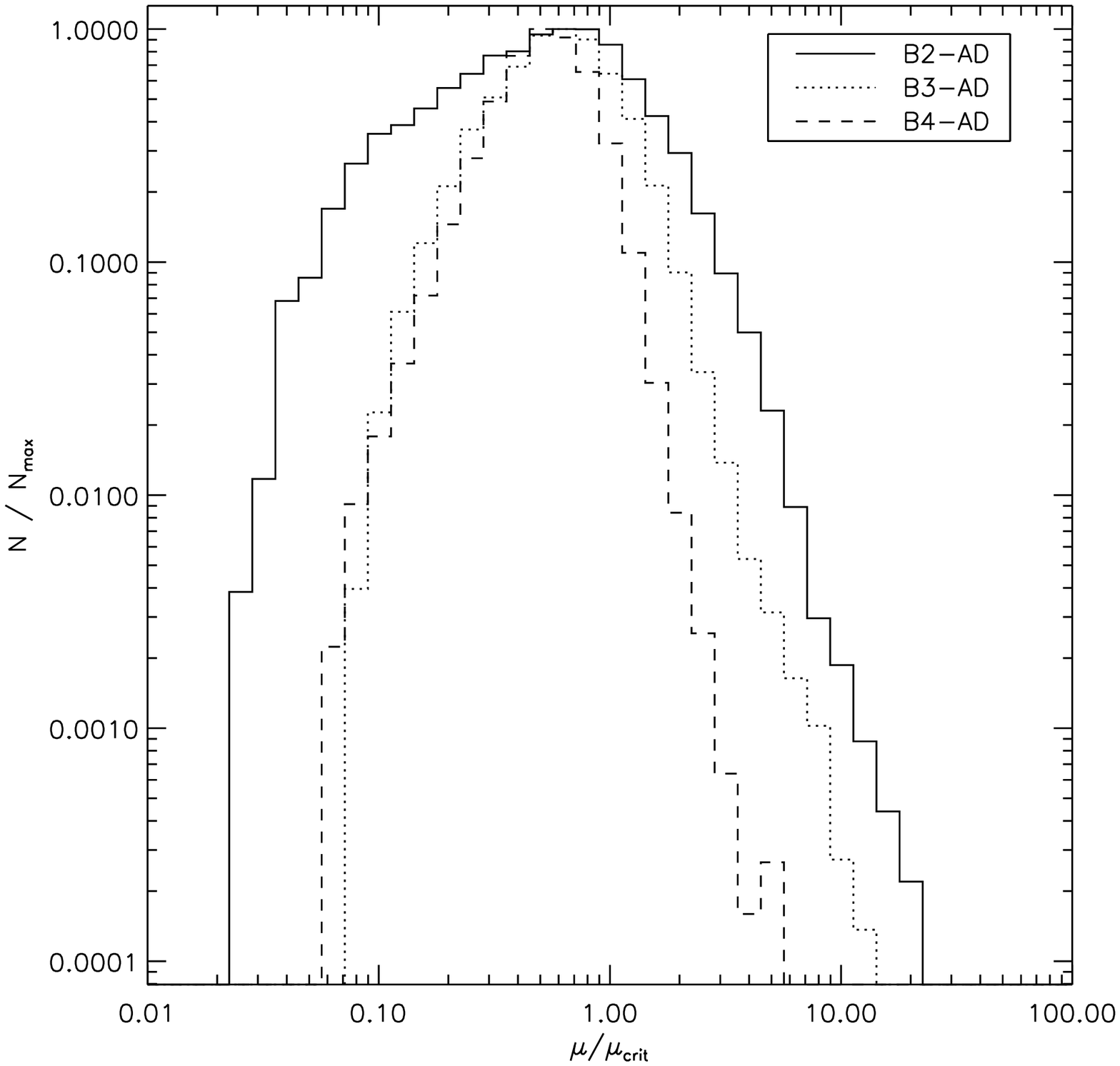}
\includegraphics[width=0.45\hsize]{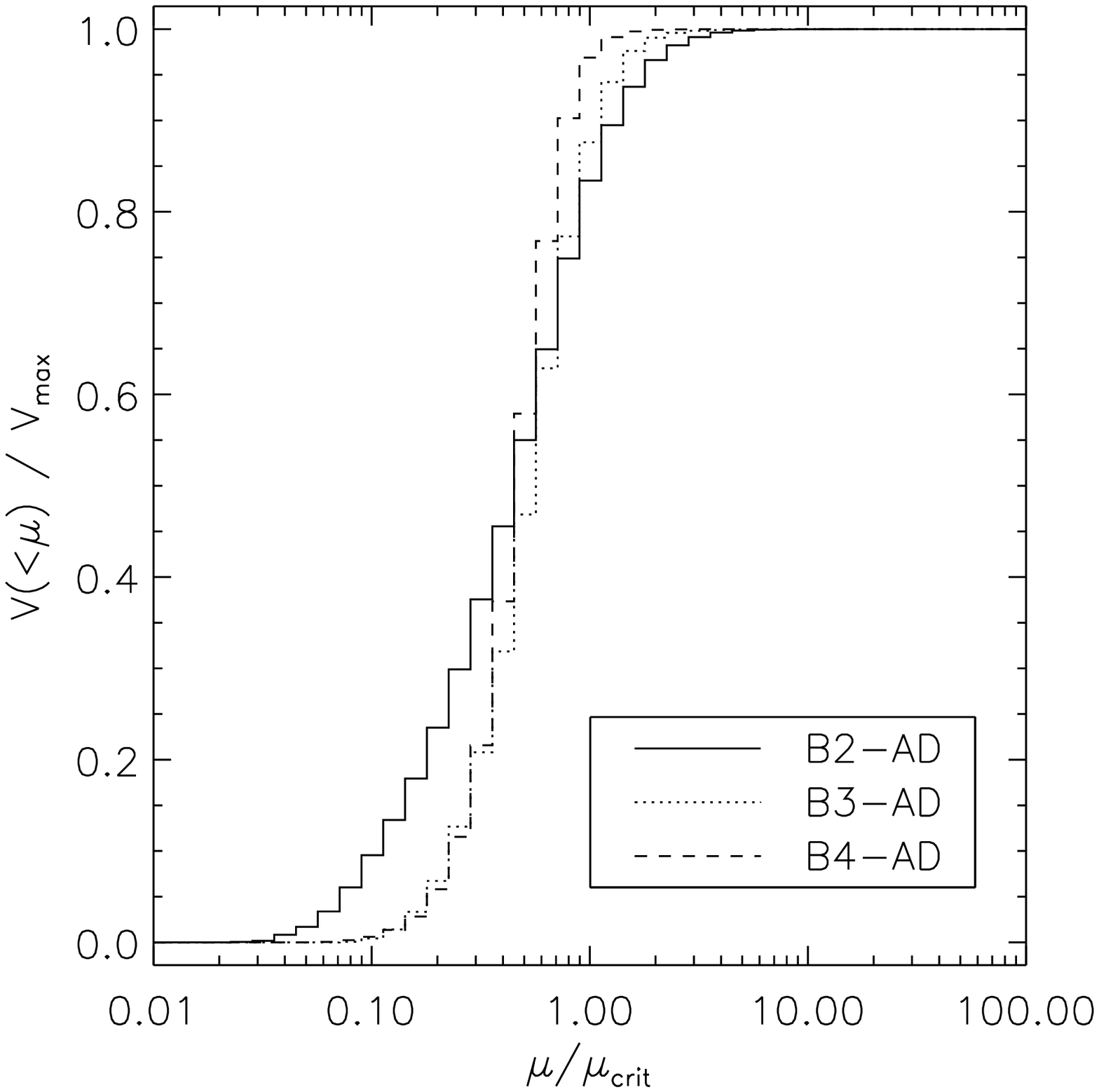}
\includegraphics[width=0.45\hsize]{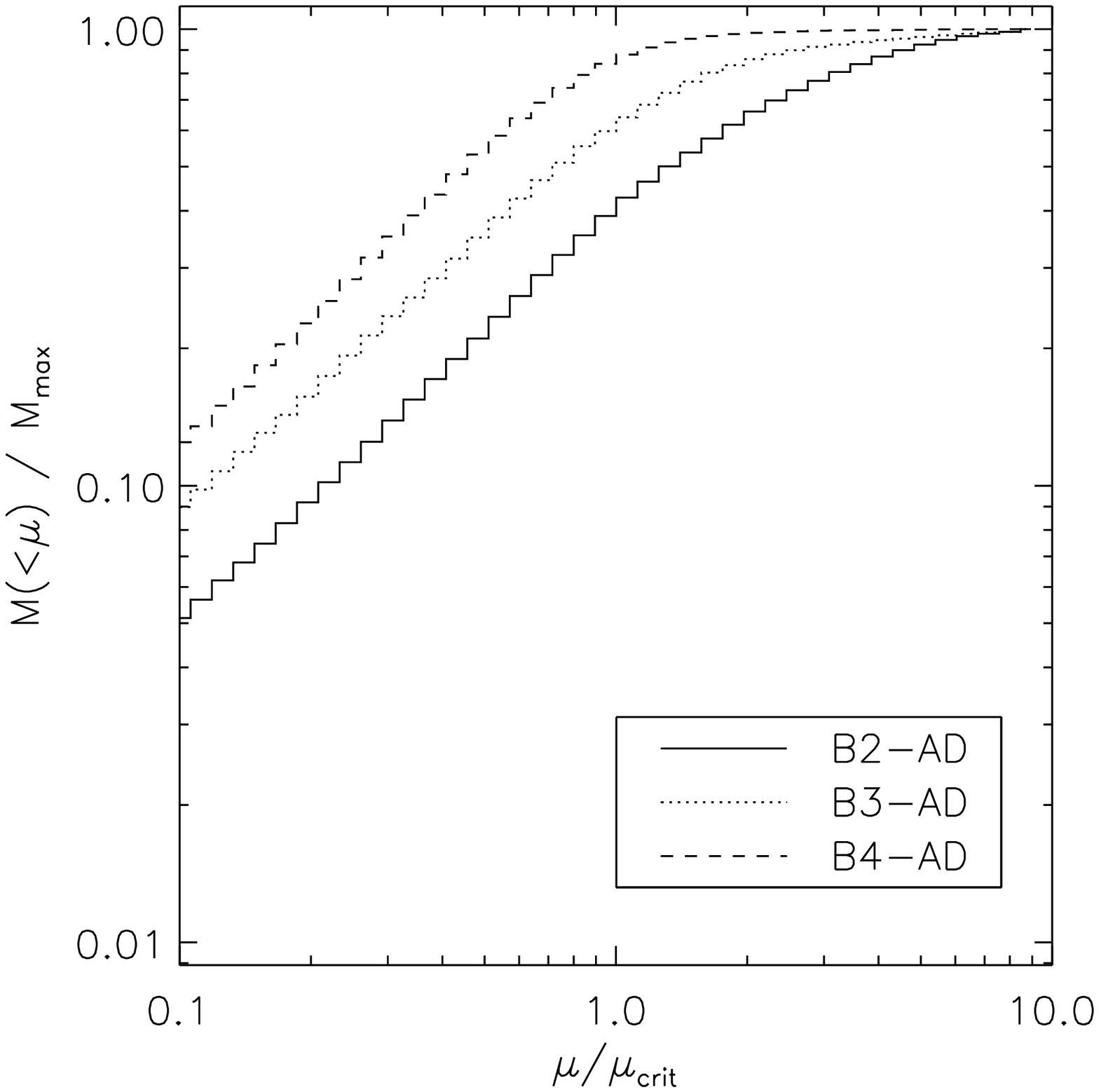}
\includegraphics[width=0.45\hsize]{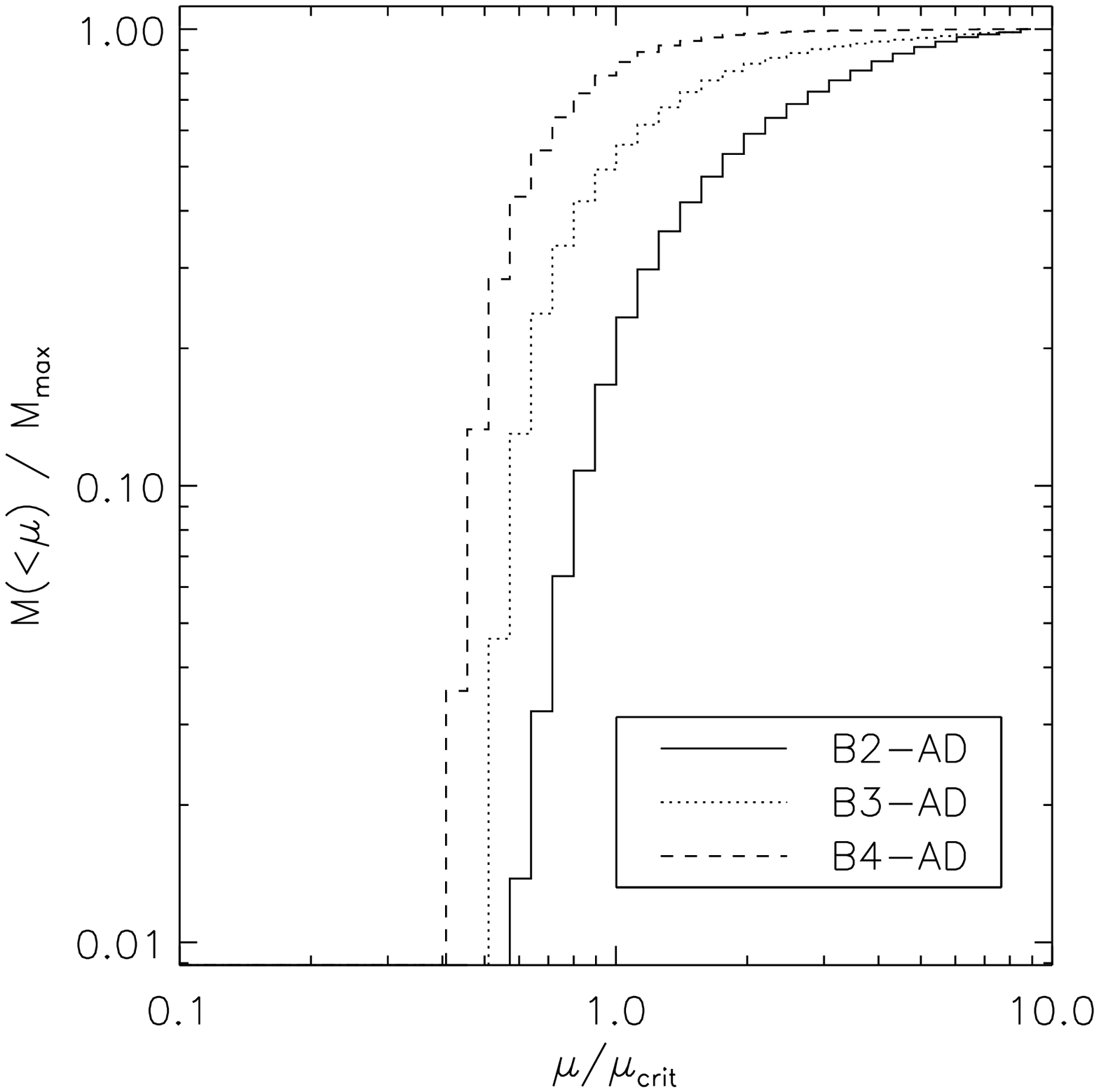}
\caption{{\it Top left:} Histograms of the M2FR, estimated through the
projection method, for runs B2-AD ({\it solid line}), B3-AD
({\it dotted line}), and B4-AD ({\it dashed line}), at $t=20$ Myr in all
cases. {\it Top right:} Cumulative probability distributions for the same
three runs, with the same line coding. {\it Bottom left} Density-weighted
probability distribution, giving the fraction of mass below the
indicated value of $\mu$. {\it Bottom right} Same as the bottom left
panel but for high-column density LOSs ($N > 10^{21} \psc$) only.}
\label{fig:mu_histo_proj}
\end{figure*}

Returning to Fig.\ \ref{fig:mu_images}, and focusing on the {\it
top row} of images, which show the variation of the M2FR's spatial
distribution as a function of the magnetic field strength, several
points are worth noting. First, as mentioned above, the M2FR is
seen to be highly inhomogeneous in all three runs. This is also
illustrated in Fig.\ \ref{fig:mu_histo_proj}, whose {\it top left panel}
shows histograms of the M2FR in the three runs at the same time as that
shown in Fig.\ \ref{fig:mu_images}. The {\it top right} and {\it bottom
left} panels of Fig.\
\ref{fig:mu_histo_proj} show the corresponding cumulative distributions,
respectively weighted by volume and by mass. Finally, the {\it bottom
right} panel shows the mass-weighted cumulative distribution for
high-column density ($N > 10^{21} \psc$) LOSs only. From these figures,
it is seen that $\mu$ fluctuates by over one order of magnitude in the
subcritical runs, and by two in the supercritical one.  Part of this
variability, especially the highest values of $\mu$, may be an artifact
of the measurement procedure, as discussed in Sec.\
\ref{sec:consid_M2FR}.  Nevertheless, significant actual fluctuations of
the M2FR on the plane are expected, since the cloud is turbulent and
clumpy, due to the combined action of the thermal, Kelvin-Helmholz and
nonlinear thin-shell \citep[NTSI,][]{Vishniac94} instabilities
\citep{VS_etal06, Heitsch_etal06}. In the ideal MHD case, segments of
magnetic flux tubes must have M2FRs smaller than that of the whole tube
at all times. However, in the presence of diffusion (numerical and/or
ambipolar), Lagrangian regions where the density is large enough can
lose magnetic flux and reach M2FR values larger than the initial value
for the tube. Thus, local clumps may become magnetically supercritical
even within the globally subcritical simulations, as in the low-mass
mode of the ``standard model'' of magnetically regulated SF
\citep{SAL87, Mousch91}, and turbulent extensions of it
\citep{NL05}. According to the {\it top right} and {\it and bottom left}
panels of Fig.\ \ref{fig:mu_histo_proj}, the fraction of the volume
(resp.\ mass) that is in locally supercritical regions increases from
$\sim 3$\% (resp.\ $\sim 10$\%) in run B4-AD to $\sim 17$\% (resp.\
$\sim 58$\%) in run B2-AD. In summary, there exist plenty of mechanisms
that contribute to the development of a highly inhomogeneous spatial
distribution of the M2FR, both in the ideal MHD and in the diffusive
cases. This is consistent with recent observational determinations
suggesting that the magnetic field strength is randomly distributed in
MCs, with only its maximum values scaling as a power law of the density
\citep{Crutcher_etal10}.

Second, we note from the {\it top left panel} of Fig.\
\ref{fig:mu_histo_proj} that the M2FR histogram for the supercritical
run B2-AD is wider than the histograms of either of the subcritical
runs, having a larger fraction of {\it both} sub- and supercritical
LOSs. This can be understood as a consequence of the combined
action of AD enhanced by turbulence and mass conservation. For weaker
magnetic fields, the Alfv\'enic Mach number is larger, implying larger
density fluctuations, in which AD is enhanced \citep{FA02,
Heitsch_etal04, LN04}, thus allowing the formation of more strongly
supercritical clumps, which are denser and contain more mass. In turn, this
implies a stronger evacuation of the remaining regions, which are those
left with smaller masses, and therefore with lower values of the
M2FR. 

\begin{figure*}
\includegraphics[width=0.9\hsize]{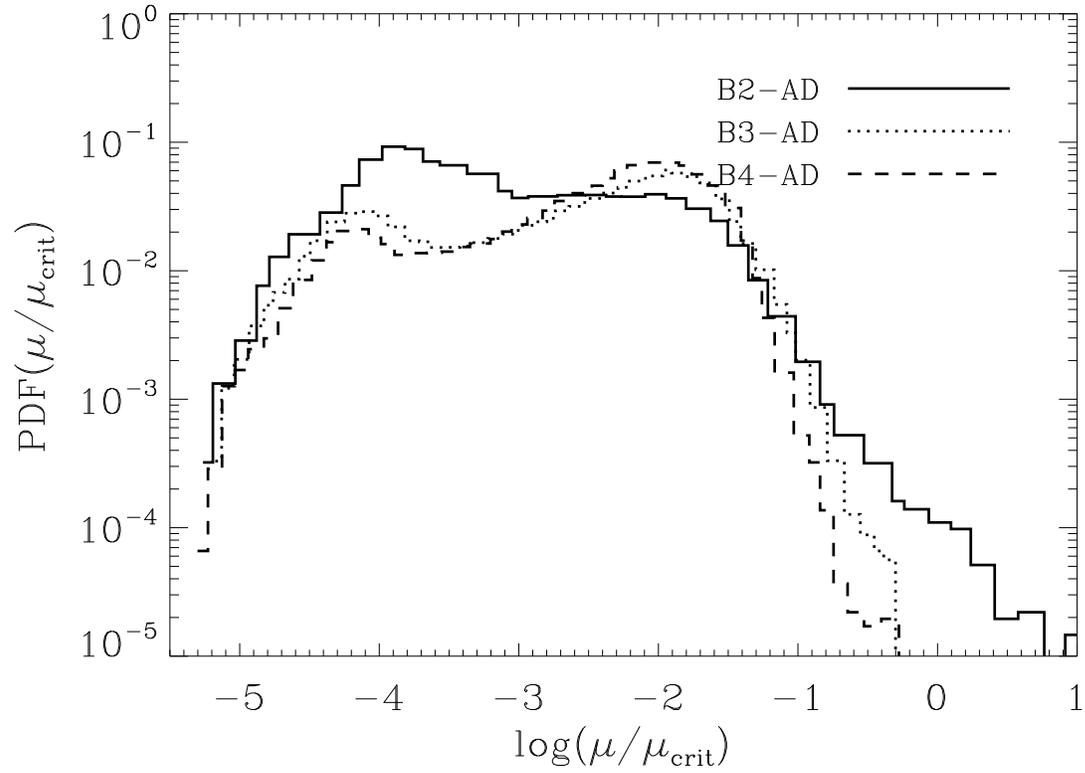}
\caption{{\it Left:} Histograms of the M2FR at $t=20$ Myr, estimated
through the 
``local method'', using eq.\ (\ref{eq:mu_local}) for each cell of the
middle plane, $x=0$, of the three simulations, B2-AD ({\it solid line}),
B3-AD ({\it dotted line}) and B4-AD ({\it dashed line}). These
histograms can be compared to those using the projection method, shown
in the {\it left panel} of Fig.\ \ref{fig:mu_histo_proj}.}
\label{fig:mu_histo_loc}
\end{figure*}

Third, from the images in Fig.\ \ref{fig:mu_images}, in which the dots
indicate the positions of the sink particles, we see that not all of the
regions that appear supercritical according to the projection method
proceed to form stars. This may be either because they are truly
supercritical albeit locally Jeans-stable, or because they are actually
magnetically subcritical, and only appear supercritical due to the
projection effect. Therefore, it is important to determine the degree to
which the M2FR is overestimated by the projection method. We defer a
detailed energy-balance study of the clumps and cores for a future
paper, but here we can take a first step towards addressing this problem
by comparing the maps and histograms of the M2FR obtained with the
projection method to those obtained with the ``local method'' (cf.\
Sec.\ \ref{sec:consid_M2FR}).  Figure\ \ref{fig:mu_histo_loc} shows
histograms of the M2FR using this method for the middle plane of each
simulation at $t=20$ Myr. Here we do not integrate over any LOS in order
to show the 
largest excursions that the M2FR can exhibit in local cells. As
mentioned in Sec.\ \ref{sec:consid_M2FR}, the local method gives lower
limits to the actual M2FR in the flux tube to which the local cell
belongs.

The $\mu$-histograms using the local method exhibit a number of
interesting features. First, it is seen that no supercritical cells are
seen in both of the subcritical runs. At least for run B3-AD this is
necessarily an underestimation of the actual M2FR, since sink formation
has already occurred in this run at the time at which the histograms are
made ($t=20$ Myr). Second, the supercritical run B2-AD exhibits a small
but finite fraction of supercritical cells, even with this
M2FR-underestimating method, indicative of the abundance of
supercritical regions in this case. Third, the histograms are seen to
peak at $\mu/\mucrit \sim 10^{-2}$, while those obtained with the
projection method peak at $\mu/\mucrit \sim 0.6$. Thus, the difference
between the two methods is so large that the true distribution remains
relatively unconstrained between them.

A final criterion that can be used to determine the accuracy of the
observational-like projection method is to compare the supercritical
mass fraction obtained with this method with the SFE, which is in fact a
measure of 
the mass fraction that has become simultaneously Jeans-unstable and
magnetically supercritical over the cloud's history. Unfortunately, in
principle this relationship is not trivial, because the supercritical
mass fraction is an instantaneous quantity in the simulation, while the
stellar mass in the cloud is a time-integrated quantity. However, if an
approximately stationary state is established in which fresh gas is
continuously added to the clouds by accretion from the WNM, replenishing
the gas used up to form stars \citep{VS_etal10}, then the (stationary)
SFE may be considered as a lower limit to the (stationary) supercritical
mass fraction. The notion that the SFE is a lower limit of the
supercritical mass fraction allows for the possibility that efficiency
of conversion of gas to stars is still less than unity even within the
supercritical, Jeans-unstable gas. 

\begin{figure*}
\includegraphics[width=0.9\hsize]{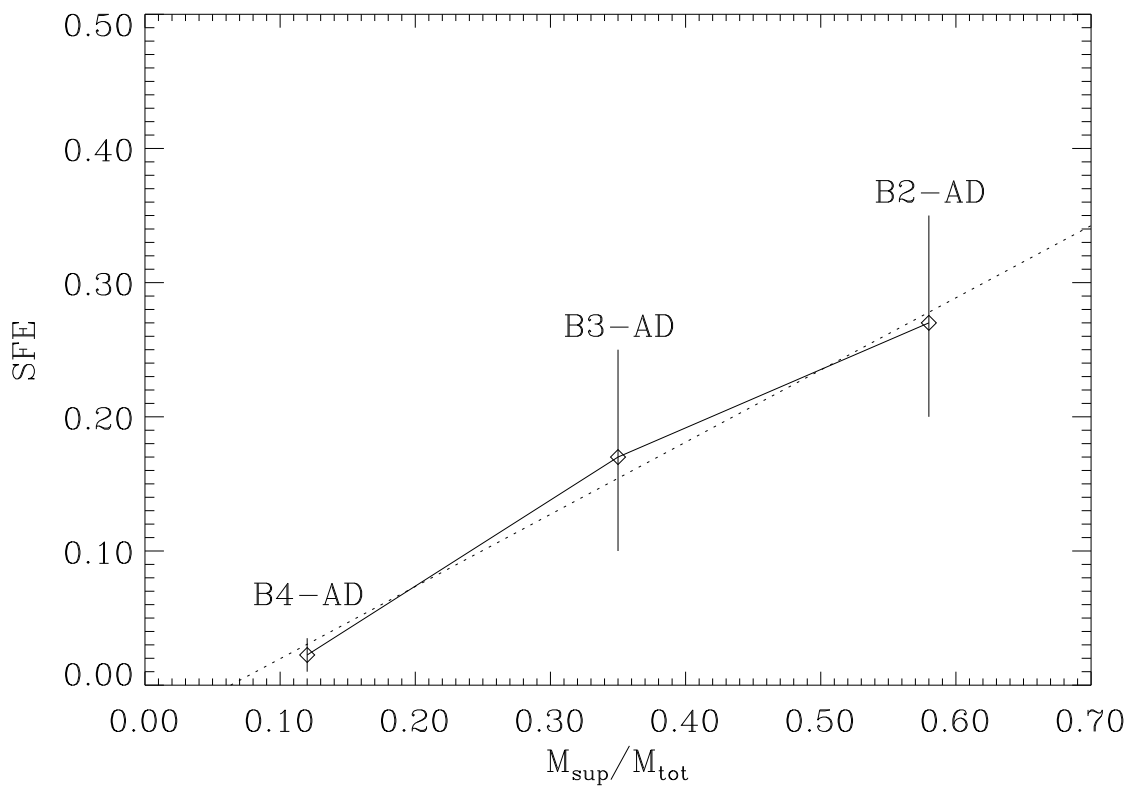}
\caption{SFE {\it versus} the supercritical mass fraction in runs B2-AD,
B3-AD and B4-AD, with the SFE read off from the {\it bottom right} panel
of Fig.\ \ref{fig:sink_gas_SFE_evol} and the supercritical mass fraction
read off from the {\it bottom left} panel of Fig.\
\ref{fig:mu_histo_proj}. The plotted value of the SFE is the mean
between the extremes taken by the SFE over the time interval after which
the initial rapid growth has ended ($18 \la t \le 26$ Myr for B2-AD, $25
\la t \le 36$ Myr for B3-AD, and $30 \la t \le 58$ Myr for B4-AD), and
the error bars denote these 
extremes. The dotted line indicates a least squares fit to the data
points, given by eq.\ (\ref{eq:sfe_vs_msup}).}
\label{fig:sfe_vs_sup}
\end{figure*}

With this in mind, we plot in Fig.\ \ref{fig:sfe_vs_sup} the SFE versus
the supercritical mass fraction in the three runs B2-AD, B3-AD and
B4-AD, as read off from Figs.\ \ref{fig:sink_gas_SFE_evol} ({\it bottom
right} panel) and \ref{fig:mu_histo_proj} ({\it bottom left} panel). The
plotted value of the SFE is the mean between the extremes taken by the
SFE over the time interval after which the initial rapid growth has
ended, and the error bars denote these extremes. The dotted line
indicates a least squares fit to the data points, given by
\beq
{\rm SFE} \approx -0.034 + 0.54 \left(\frac{\Msup}{M}\right).
\label{eq:sfe_vs_msup}
\eeq
Of course, this fit is totally empirical, and is provided only as a
guideline for the trend of the SFE with the supercritical mass
fraction. In particular, it is meaningless below the value of
the supercritical mass fraction that produces an SFE of zero.

We observe that, in all three cases, the SFE is a few to several times
smaller than the supercritical mass fraction. Under the assumption of
stationarity, this then suggests that the supercritical mass fraction
obtained through the projection method does not differ from the real one
by more than factors of a few, on average.

A final point worth noting is that, as shown in the {\it bottom right}
panel of Fig.\ \ref{fig:mu_histo_proj}, the supercritical mass fraction
in the high-column density gas is significantly larger in all three runs
than the supercritical mass fraction for gas at all column densities
({\it bottom left} panel). This reinforces the scenario that the M2FR is
determined mainly by the column density acquired by the individual
regions in the clouds by accretion of gas along field lines, and less
importantly by the local effect of ambipolar (and/or numerical)
diffusion. 

We conclude from this section that the projection method 
gives estimates of the M2FR that are within less than a factor of a few
from the actual distribution, and that the dominant mechanism that
determines the local M2FR is the accumulation of gas along field lines.

\subsection{Evolution of the global M2FR} \label{sec:M2FR_evol}

\begin{figure*}
\includegraphics[width=0.9\hsize]{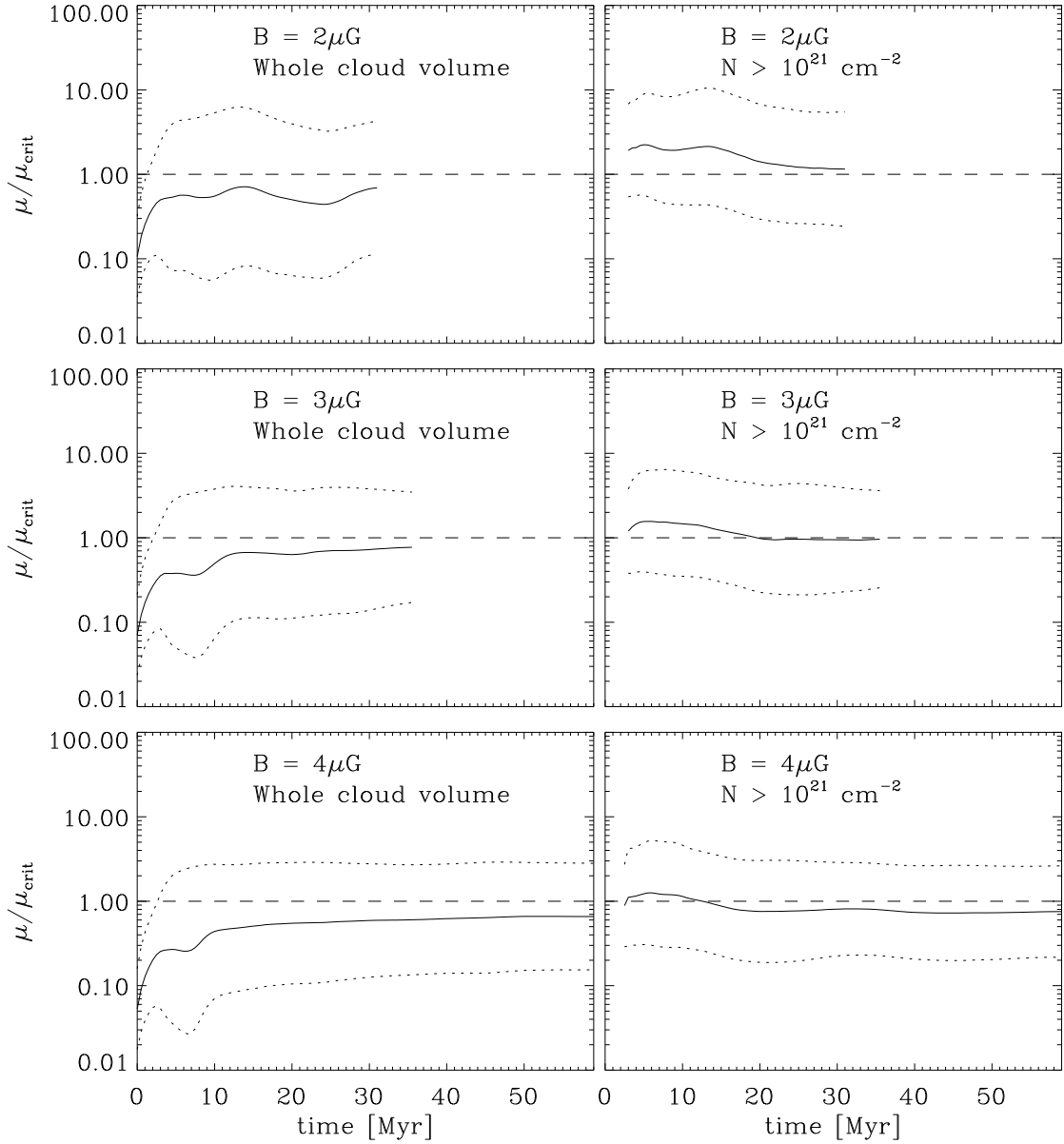}
\caption{Evolution of the M2FR $\mu$, normalized to the critical value,
for runs B2-AD ({\it top}), B3-AD ({\it middle}) and B4-Ad ({\it
bottom}), computed using the projection method over the same circular
region as in Fig.\ \ref{fig:mu_histo_proj}. The {\it solid} lines show
the mean value of $\mu/\mucrit$, and the {\it dotted} lines delimit the
$3\sigma$ range of $\mu$, where the mean and the standard deviation are
calculated for $\log \mu$. The {\it left} panels show these quantities
computed for all lines of sight parallel to the axis of a cylindrical
volume of length and diameter both equal to 20 pc, centered in the
centre of the numerical box. The {\it right} panels show the same
quantities computed only for lines of sight having column densities $N >
10^{21} \psc$. In all cases, the lines of sight extend over the interval
$-10 < x < 10$ pc.}
\label{fig:mu_evol_8_10_11}
\end{figure*}

We proceed now to discuss the evolution of the mean M2FR of the clouds
in the simulations, and its range of variability. Figure
\ref{fig:mu_evol_8_10_11} shows the evolution of the mean and the $3
\sigma$ range of the normalized M2FR, $\mu/\mucrit$, for runs B2-AD
({\it top} panels), B3-AD ({\it middle} panels), and B4-AD ({\it bottom}
panels). The computation of the M2FR is performed using the projection
method with the same path length and over the same circular region as
those used for the histograms of Fig.\ \ref{fig:mu_histo_proj}. The {\it
left} panels of Fig.\ \ref{fig:mu_evol_8_10_11} show the mean and $3
\sigma$ range computed for the set of all lines of sight contained in
the circular region, while the {\it right} panels show these quantities
computed only for the set of lines of sight for which the column density
is larger than $10^{21} \psc$.

These plots illustrate the fact, discussed in Sec.\ \ref{sec:general},
that the M2FR of a cloud is an evolving quantity, which first increases
as the cloud gathers material from the WNM inflows that assemble it. At
the inflow speed of $13.9 \kms$, the 112-pc-long inflows are entirely
incorporated into the cloud in approximately 8 Myr, which indeed
corresponds to the time scale at which the M2FR of all three clouds is
seen to have reached a roughly stationary value. This value is seen to
be larger for weaker mean field, as expected. However, a second increase
in the M2FR is seen to occur after this first stabilization. This can be
attributed to the fact that the inflows drag part of the surrounding,
initially static WNM along with them, as they leave a partial vacuum behind
them. This dragged material flows at lower speeds, and reaches the cloud
at later times \citep[for further discussion, see][]{VS_etal07}.

It is interesting that $\langle \mu/\mucrit \rangle$ for the set of all
lines of sight is smaller than unity at all times for all three runs,
even the supercritical one, B2-AD. This is likely a consequence of mass
conservation again (cf.\ Sec.\ \ref{sec:M2FR_distr}). Because these
statistics are weighted by area over the circular region, the
supercritical regions, which are denser, occupy a smaller fraction of
the surface area of the cloud, and therefore the area-weighted average
M2FR is smaller than unity in all cases. However, the averages for the
high-column density LOSs, shown in the {\it right panels} of Fig.\
\ref{fig:mu_evol_8_10_11} are larger than unity for all times in run
B2-AD, and for nearly 20 Myr in run B3-AD. Instead, for run B4-AD, even
this high-$N$ average barely reaches values larger than unity, and only
over less than 10 Myr. Finally, in all cases, the high-$N$ average M2FR
decreases at late times, a phenomenon that can be attributed to the
high-M2FR gas consumption by star formation.

In general, we conclude from Sections \ref{sec:M2FR_distr} and
\ref{sec:M2FR_evol} that the M2FR is a time-dependent function of
time as a cloud is built up by converging WNM streams, whose average
generally increases with time, although in the high-$N$ regions the
average later decays due to consumption by SF. Spatially, the M2FR
exhibits large fluctuations, whose 3-$\sigma$ range spans over one
order of magnitude even in the strong-field cases. It is the
high-M2FR tail of the distribution that is responsible for star formation.

\subsection{Buoyancy of subcritical regions} \label{sec:buoyancy}

\begin{figure*}
\includegraphics[height=0.31\hsize]{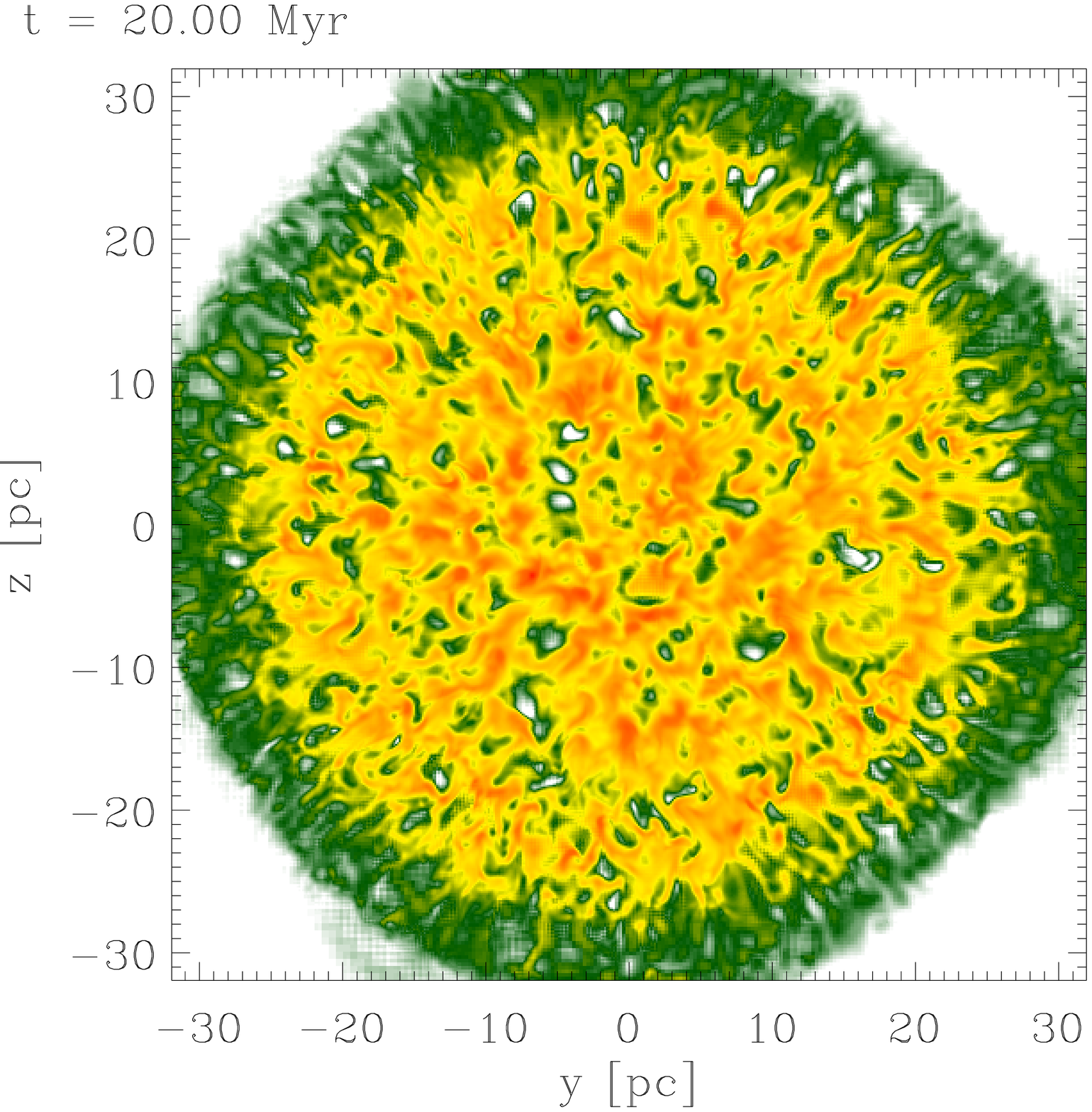}
\includegraphics[height=0.31\hsize]{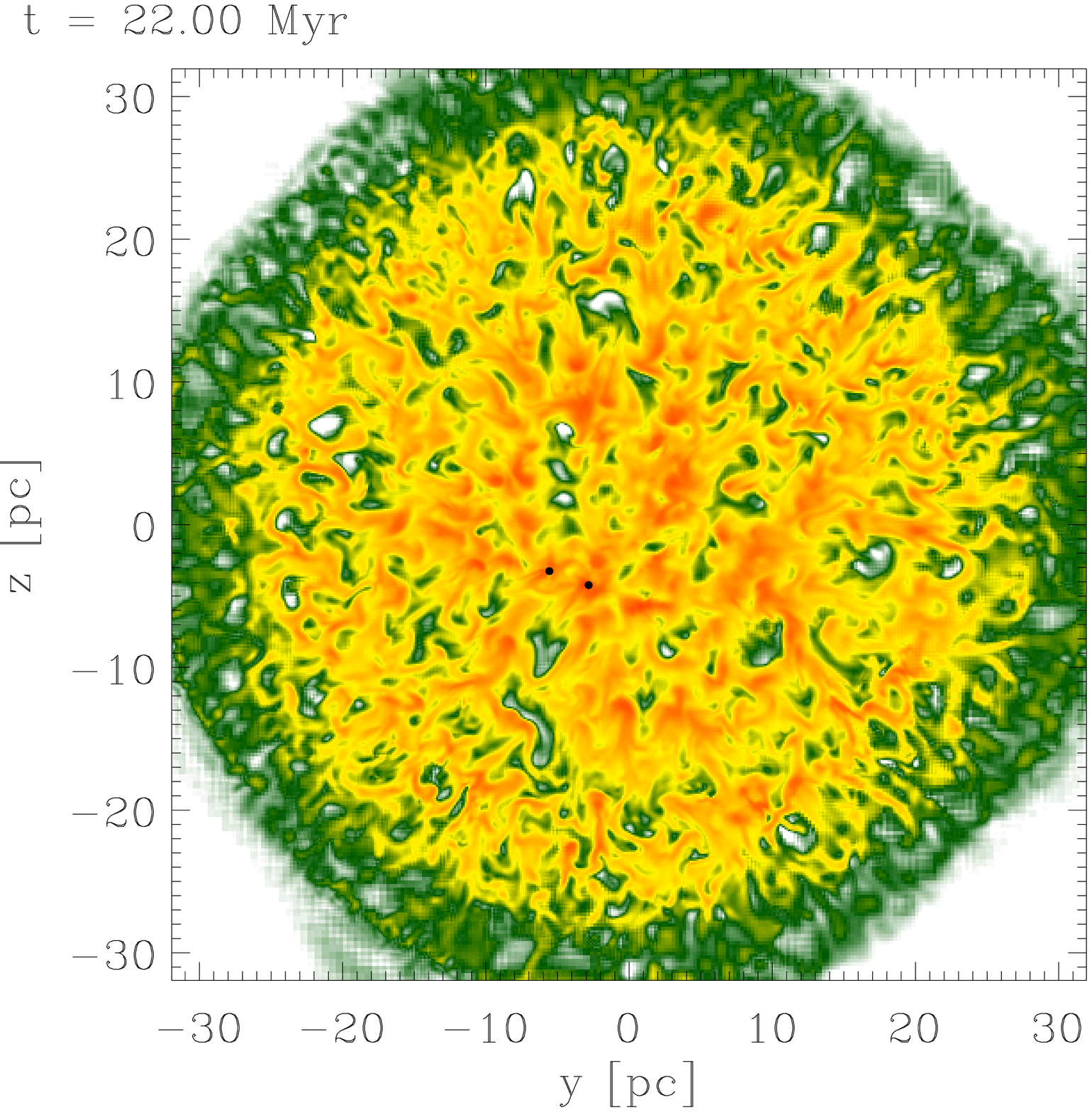}
\includegraphics[height=0.31\hsize]{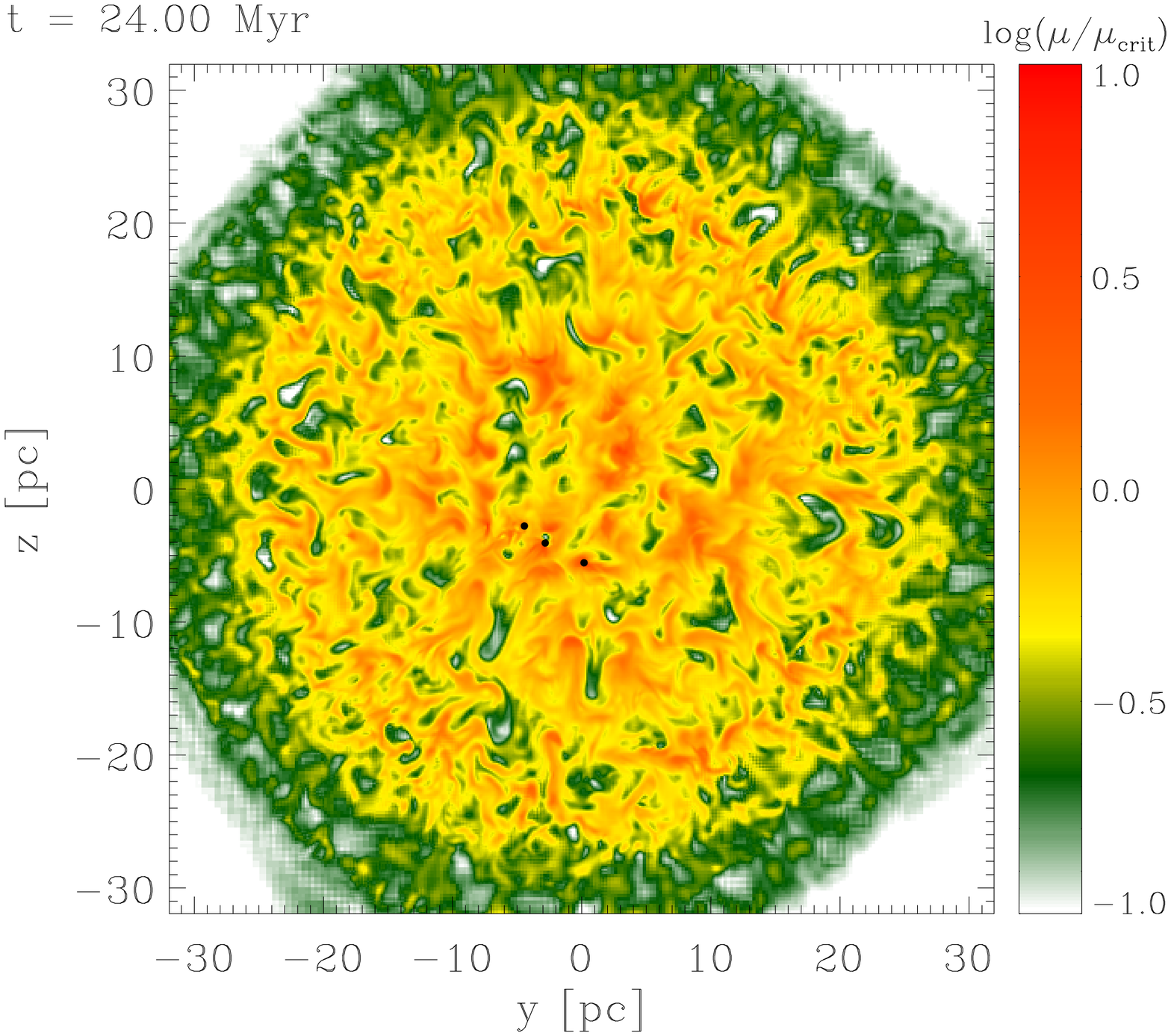}
\caption{Snapshots of the mass-to-flux ratio $\mu$, normalized to the 
critical value, for run B4-AD at times $t=20$ ({\it left panel}), 22
({\it middle panel}), and 24 ({\it right panel}) Myr. Note the evolution
of the subcritical regions, shown in green, which develop cometary
shapes, pointing outwards from the centre of the image.}
\label{fig:buoyancy}
\end{figure*}

An unexpected feature we have observed in these simulations is that the
subcritical and supercritical regions do not maintain their relative
positions fixed throughout the evolution of the simulations. Instead,
the subcritical regions exhibit ``buoyancy'', so that they tend to
separate themselves from the global contraction of the clouds, even in
the globally subcritical cases that rebound at later times. This is most
clearly seen in animations of the simulations, but can be seen in Fig.\
\ref{fig:buoyancy}, where we show that, as time proceeds, the
subcritical regions become segregated from the supercritical ones,
moving outwards, and developing cometary shapes, with their heads
pointing outwards as well.

This behavior can be described as a macroscopic-scale analogue of the
very process of AD. In the latter, the neutrals sink into the
gravitational potential well, percolating through the ions, which remain
attached to the magnetic field lines. In our case, the supercritical
regions sink into the potential well, while the subcritical ones remain
in the outer parts of the well, being held up by the magnetic
tension. The process is also reminiscent of the interchange mode of the
Parker instability \citep{HC87}.

\section{Summary and Conclusions} \label{sec:conclusion}

In this paper we have presented a study of the formation and evolution
of GMCs by the convergence of WNM streams, or ``inflows'', in the
presence of magnetic fields and AD. As described by many groups
\citep[see the review by][and references therein]{VS10}, this process
involves the transition of the atomic gas from the warm, diffuse phase
to the cold, dense one, allowing the fresh cold gas to quickly become
self-gravitating and begin to contract. 

We first reviewed the general evolution of the gas and the M2FR expected
in this type of systems, noting that, as originally pointed out by
\citet{HBB01}, the mass-to-flux ratio (M2FR) of the cloud is expected to
increase in time, so that the cloud becomes molecular, self-gravitating,
and magnetically supercritical at roughly the same time, provided that
there is enough mass in the converging streams to render them
supercritical. This condition requires that, for solar neighborhood
conditions, the inflows extend beyond the accumulation length given by
eq.\ (\ref{eq:acc_length}). Flows that do not extend to such distances
are expected to form subcritical clouds which, however, may be
predominantly atomic. This suggests that, in particular, the converging
flows induced by the spiral-arm potential wells, which have typical size
scales $\sim 1$ kpc, will in general induce the formation of
magnetically supercritical molecular clouds. On the other hand,
converging flows induced by smaller-scale inflows, such as supernova
shocks, or simply turbulent random motions in the gas, may lead to the
formation of subcritical, partially atomic clouds.

We then discussed the difficulties inherent to measuring the M2FR, even
under controlled conditions such as those of the simulations. Other than
measuring the M2FR along magnetic flux tubes, which is impossible to perform
observationally, and perhaps even numerically, we considered two methods for
measuring the M2FR: the ``projection'' method, which mimicks the
observational procedure of measuring the ratio of column density to
field strength along the line of sight (LOS), and which gives an upper
limit to the actual M2FR, and the ``local'' method, which simply
measures the mass and magnetic field in a grid cell in the simulation,
and estimates the M2FR as given by eq.\ (\ref{eq:mu_local}),
giving a lower limit to the M2FR.

We next studied the evolution of the M2FR and the star-forming
properties of clouds formed by both sub- and supercritical inflows. We
concluded that in our simulations, the effect of numerical diffusion is
at a comparable level to that of AD. We found that the subcritical cases
do undergo an initial phase of contraction, followed by a re-expansion,
settling into an oscillatory regime. The supercritical case, on the
other hand, proceeds directly to collapse, as expected. All cases form
stars, although at greatly different rates, producing what appears more
a continuum of star-forming regimes as the mean magnetic field strength
is varied, rather than a bimodal regime of high SFR in supercritical
cases and low SFR in subcritical ones, as was the case in the
``standard'' model of magnetically regulated SF \citep{SAL87,
Mousch91}. In particular, the marginally subcritical case B3-AD, through
the action of diffusion, reached SF efficiencies (SFEs) comparable to
those of the supercritical case B2-AD. The onset of SF is delayed by up
to 15 Myr in the most strongly magnetized cases we studied, although
this delay is reduced by a few to several Myr when AD is included (when
it is not, all SF activity is due to numerical diffusion). The SFEs
observed in our simulations range from $\sim 35$\% for run B2-AD to $\sim 3$\%
for run B4-AD. However, since stellar feedback, which would reduce the
SFE even further, is not included in these simulations, it is likely
that the efficiency of run B4-AD is actually too low in comparison with
observed values.

We then investigated the spatial and statistical distribution of the
M2FR, finding that this is a highly fluctuating quantity. The
fragmentation of the cloud by the combined action of thermal, nonlinear
thin-shell, and gravitational instabilities leads to the formation of
clumps of high density and high M2FR, and of low-density, low-M2FR
patches. The fluctuations in the M2FR we observed span between
one-and-a-half and two orders of magnitude, the distribution being wider
for weaker magnetizations. These results are qualitatively consistent
with recent observational determinations suggesting that the magnetic
field strength in MCs is strongly fluctuating \citep{Crutcher_etal10}.

Next, we discussed the evolution of the mean M2FR and its 3-$\sigma$
range in the various simulations, finding that in general it initially
increases as expected by the assembly process of the cloud, to later
reach a roughly stationary regime. The M2FR of the high-column-density
LOSs, on the other hand, tends to decrease at later times, due to the
consumption of gas in this regime by SF.

Finally, we reported the occurrence of an unexpected effect: the
buoyancy of the low-M2FR regions with respect to the high-M2FR ones,
through a process that appears as the macroscopic-scale analogue of AD:
the high-M2FR regions sink deeper into the potential well of the cloud,
while the low-M2FR ones remain supported by the field in the outer parts
of the cloud, so that the clouds evolve towards a segregated state with
low M2FR in their periphery and high-M2FR towards their centre, even on
scales much larger, and densities much lower than, those directly
affected by AD. The process also bears resemblance with the intercheange
mode of the Parker instability.

\section*{Acknowledgments}

The FLASH code was developed in part by the DOE-supported Alliances
Center for Astrophysical Thermonuclear Flashes (ASC) at the University
of Chicago. Our simulations were carried out at the Leibniz
Rechenzentrum, Garching and the J\"ulich Supercomputing Centre
(JSC). We gratefully acknowledge the following sources of financial
support: CONACYT (M\'exico) grants U47366-F and 102488 to E.V.-S.;
Emmy-Noether (DFG) grant BA 3706 to R.B.; International Collaboration II
(grant P-LS-SPII/18) program of the Landesstiftung Baden-W\"{u}rttemberg,
subsidies from the Deutsche Forschungsgemeinschaft under grants no. KL
1358/1, KL 1358/4, KL 1359/5, KL 1358/10, and KL 1358/11, and a
Frontier grant of Heidelberg University sponsored by the German
Excellence Initiative to R.S.K.


\end{document}